\documentclass{aastex63}

\usepackage{enumitem}
\usepackage{amsmath}
\newcommand\numberthis{\addtocounter{equation}{1}\tag{\theequation}}
\newcommand{\RNum}[1]{\uppercase\expandafter{\romannumeral #1\relax}}

\newcommand{\lya}{Ly$\alpha$}
\newcommand{\Rearth}{R$_{\oplus}$}
\newcommand{\Mearth}{M$_{\oplus}$}
\newcommand{\Rsol}{R$_{\odot}$}

\shorttitle{\sc AU Mic b's Escaping Atmosphere}
\shortauthors{Rockcliffe et al.}

\begin{document}

\title{\sc The Variable Detection of Atmospheric Escape around the young, Hot Neptune AU Mic b}

\author[0000-0003-1337-723X]{Keighley E. Rockcliffe}
\affiliation{Department of Physics and Astronomy, Dartmouth College, Hanover, NH 03755, USA}

\author[0000-0003-4150-841X]{Elisabeth R. Newton}
\affiliation{Department of Physics and Astronomy, Dartmouth College, Hanover, NH 03755, USA}

\author[0000-0002-1176-3391]{Allison Youngblood}
\affiliation{NASA Goddard Space Flight Center, Greenbelt, MD 20771, USA}

\author[0000-0002-7119-2543]{Girish M. Duvvuri}
\affiliation{Department of Astrophysical \& Planetary Sciences, University of Colorado Boulder, Boulder, CO 80303, USA}

\author[0000-0002-8864-1667]{Peter Plavchan}
\affiliation{Department of Physics and Astronomy, George Mason University, Fairfax, VA 22030, USA}

\author[0000-0002-8518-9601]{Peter Gao}
\affiliation{Earth and Planets Laboratory, Carnegie Institution for Science, 5241 Broad Branch Road, NW, Washington, DC 20015, USA}

\author[0000-0003-3654-1602]{Andrew W. Mann}
\affiliation{Department of Physics and Astronomy, University of North Carolina at Chapel Hill, Chapel Hill, NC 27599, USA}

\author[0000-0001-8014-0270]{Patrick J. Lowrance}
\affiliation{IPAC, MS 220-6, California Institute of Technology, Pasadena, CA 91125, USA}

\begin{abstract}

Photoevaporation is a potential explanation for several features within exoplanet demographics. Atmospheric escape observed in young Neptune-sized exoplanets can provide insight into and characterize which mechanisms drive this evolution and at what times they dominate. AU Mic b is one such exoplanet, slightly larger than Neptune ($4.19$ \Rearth). It closely orbits a $23$ Myr pre-Main Sequence M dwarf with a period of $8.46$ days. We obtained two visits of AU Mic b at Lyman-$\alpha$ (\lya) with {\it HST}/STIS. One flare within the first {\it HST} visit is characterized and removed from our search for a planetary transit. We present a non-detection in our first visit followed by the detection of escaping neutral hydrogen {\it ahead of the planet} in our second visit. The outflow absorbed \replaced{$\sim 50$\%}{$\sim 30$\%} of the star's \lya\ blue-wing $2.5$ hours {\it before} the planet's white-light transit. We estimate the highest velocity escaping material has a column density of $10^{13.96}$ cm$^{-2}$ and is moving $61.26$ km s$^{-1}$ away from the host star. AU Mic b's large high energy irradiation could photoionize its escaping neutral hydrogen in $44$ minutes, rendering it temporarily unobservable. Our time-variable \lya\ transit ahead of AU Mic b could also be explained by an intermediate stellar wind strength from AU Mic that shapes the escaping material into a leading tail. Future \lya\ observations of this system will confirm and characterize the unique variable nature of its \lya\ transit, which combined with modeling will tune the importance of stellar wind and photoionization.

\end{abstract}


\section{Introduction} \label{sec:intro}

Thousands of exoplanets within the Milky Way have been discovered with transit photometry over the last twenty years, with significant contributions from {\it Kepler} and now {\it TESS} \footnote{\href{https://exoplanetarchive.ipac.caltech.edu/docs/counts_detail.html}{https://exoplanetarchive.ipac.caltech.edu/docs/counts\_detail.html}}. A significant feature of the exoplanet population is the ``radius gap". The gap, defined by the relative lack of planets with radii between $1.5$ and $2$ \Rearth, separates two distinct populations of exoplanets: super-Earths and sub-Neptunes \citep{2018AJ....156..264F,2017ApJ...847...29O,2017AJ....154..109F,2013ApJ...775..105O}. A leading theory is that atmospheric escape produced the gap, where a population of larger and less dense planets closely orbited their host stars and subsequently lost envelope mass over time, leaving behind cores with either little-to-no atmosphere (super-Earths) or smaller -- a few \% by mass -- atmospheres (sub-Neptunes) \citep[e.g.,][]{2014ApJ...795...65J,2013ApJ...776....2L}. On the other hand, \cite{2021ApJ...908...32L} suggest that the radius gap is a product of delayed gas accretion during exoplanet formation, and not of atmospheric escape or other evolutionary processes.

Another feature is the ``hot Neptune desert", a region in exoplanet radius--period space with a paucity of Neptune-sized planets with short orbital periods \citep[{P$_{\text{p}}\lesssim3$ days};][]{2016NatCo...711201L,2007A&A...461.1185L}. We expect the exoplanets born within the desert to be rocky core planets with large gaseous envelopes that quickly evolve out of the desert. As shown in \cite{2018MNRAS.479.5012O} and \cite{2016A&A...589A..75M}, the desert could be a consequence of atmospheric escape and orbital migration. These large, low density planets are part of the atmospheric escape story that formed the previously mentioned radius gap. These two features and their contested origins motivate continued study of atmospheric escape.

Any atmosphere can experience mass loss. The Earth loses hydrogen from its upper atmosphere at roughly $3$ kg s$^{-1}$, a negligible amount compared to the Earth's atmospheric mass and replenishment \citep{2009SciAm.300e..36C}. Earth's mass loss can be attributed to the sum influence of several atmospheric escape processes: Jeans escape, charge exchange escape, and polar wind escape. For exoplanets, hydrodynamic escape is widely thought to be the primary atmospheric escape process that formed the hot Neptune desert and radius gap. Hydrodynamic escape occurs when there is an injection of heat to the upper atmosphere of a planet (thermosphere, exosphere), causing an outflow of gas to escape the planet's gravity. The heating mechanism is one of the key uncertainties surrounding hydrodynamic escape. Photoevaporation -- hydrodynamic escape externally driven by high-energy stellar radiation -- was the unchallenged theory for years \citep{2019AREPS..47...67O}. Recently, however, core-powered mass loss -- hydrodynamic escape internally driven by radiation from a cooling planetary core -- was posed as a viable candidate \citep{2018MNRAS.476..759G}. Both schemes are able to reproduce the radius gap, and predict the same planetary core properties and similar slopes in planetary radius--period--insolation space \citep{2017ApJ...847...29O,2019MNRAS.487...24G,2021MNRAS.503.1526R}. Alternatives, like impact-driven atmospheric escape, have also been proposed \citep{2020MNRAS.491..782W}. Photoevaporation and core-powered mass loss differ in a few ways that suggest we may be able to tease them apart observationally: outflow strength, timescale, and different slopes in planetary radius--period--host mass space. \cite{2021MNRAS.508.5886R} describes the theoretical similarities and differences between the two mechanisms in more detail.

We can use observations of atmospheric escape at different system ages and host masses to narrow down which mechanism is primarily responsible for shaping the exoplanet population. Specifically, observations of atmospheric escape in young systems ($<1$ Gyr) will not only help us figure out what mechanism dominates mass loss, but will explore at what age atmospheres are lost, how a planet's environment influences mass loss, and whether mass loss rates used in demographic studies are consistent with measurements.

Atmospheric escape has been observed using Far-Ultraviolet (FUV) transmission spectra, which contain the stellar Lyman-$\alpha$ emission line (\lya) at $1215.67$ \AA. A host star's \lya\ emission has a high likelihood of interacting with any intervening neutral hydrogen, and any escaping neutral hydrogen from a transiting exoplanet's upper atmosphere will partially absorb the incident \lya\ radiation. Interstellar clouds of neutral hydrogen (ISM) along our line of sight completely absorb the central portion of the emission line unless the star is moving fast enough to Doppler-shift the \lya\ line away from its rest-wavelength. It follows that in planetary systems with low-to-moderate radial velocities -- including all young systems -- any planetary absorption signature will only be observable in the profile wings (i.e., we are only able to observe high-velocity escaping material). Although theory predicts these outflows escape at $\sim10$s of km s$^{-1}$, which would be well within the ISM absorption of \lya\ and thus unobservable, several acceleration mechanisms (e.g., stellar wind ram pressure, charge exchange, radiation pressure) have been proposed and modeled that can boost some of the planetary wind to $\sim100$s of km s$^{-1}$ \citep{2019ApJ...873...89M,2018A&A...620A.147B,2013A&A...557A.124B}.

{\it Hubble Space Telescope (HST)} Space Telescope Imaging Spectrograph (STIS) can be used to obtain high signal-to-noise observations at \lya\ at times surrounding the white-light transit of a planet. If planetary absorption at \lya\ is detected, the planet's hydrogen mass loss rate and the outflow geometry and velocity can be constrained. For hot Jupiters, the inferred mass loss rates are low relative to their planetary masses, so atmospheric escape has a negligible effect on the evolution of large Jupiter-massed planets \citep[e.g.,][]{2012A&A...543L...4L,2003Natur.422..143V}. Mass loss is thought to be more significant for Neptune-sized planets. There are four close-orbiting Neptunes with detections of escaping neutral hydrogen using \lya\ transmission spectroscopy: Gl 436b \citep{2019A&A...629A..47D,2017A&A...605L...7L,2015Natur.522..459E,2014ApJ...786..132K}, GJ 3470b \citep{2018A&A...620A.147B}, HD 63433c \citep{2022AJ....163...68Z}, and HAT-P-11b \citep{2022NatAs...6..141B}. In these Neptunes, \lya\ attenuation in the profile's blue-wing extends beyond the duration of the white-light transit, indicating that neutral hydrogen is escaping and traveling towards the observer in a large cometary tail-like cloud.

\begin{figure*}
    \centering
    \includegraphics[width=0.9\textwidth]{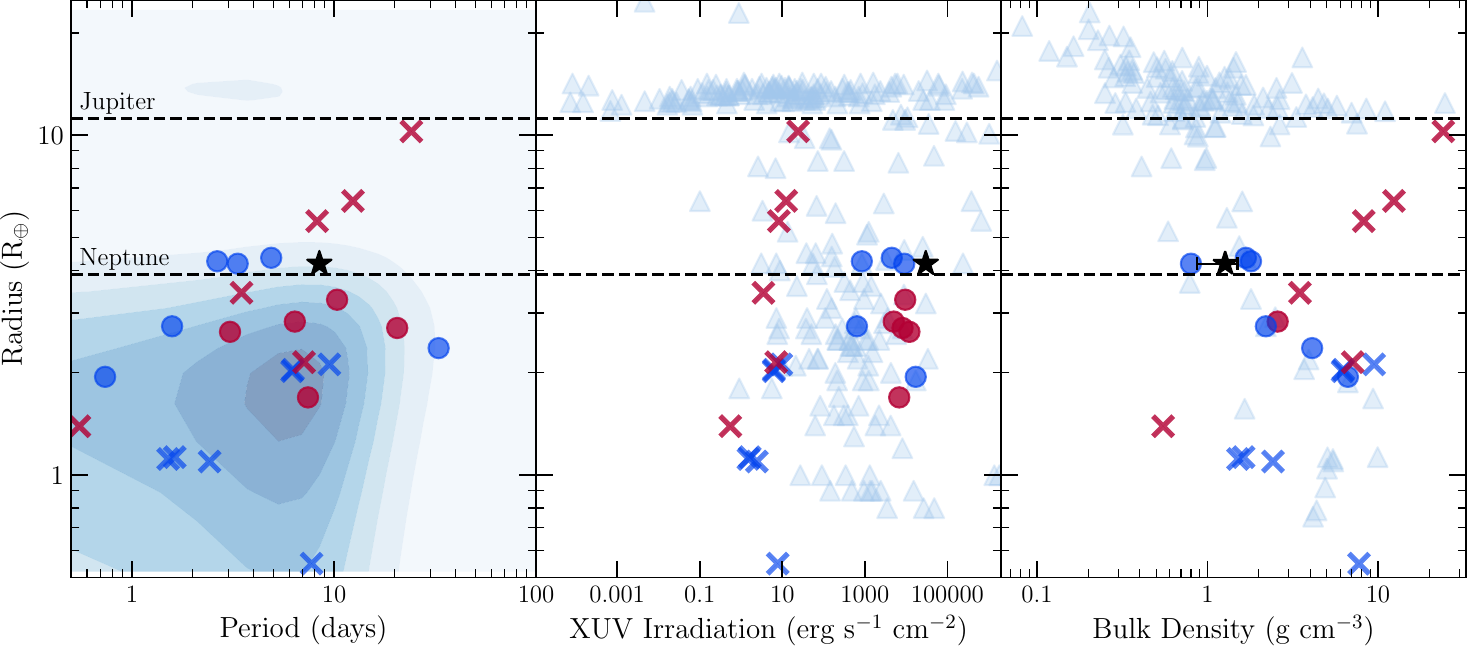}
    \caption{A panel of exoplanet demographics information from the NASA Exoplanet Archive as of August 2022. Non-specified, field-age planets are represented by the light blue contours and light blue triangles. Specified planets have been searched for atmospheric escape. Filled circles indicate at least one published detection of hydrogen and/or helium escape. The x's indicate hydrogen/helium escape non-detections. Among the detections/non-detections, field-age planets are dark blue, whereas young planets are red. The black star represents AU Mic b, with its range in bulk density showing its spread in mass measurements. The leftmost panel shows the distribution of exoplanets in radius and period. The center panel is exoplanet radius versus its X-ray to ultraviolet irradiation \citep{2022A&A...661A..23F}. The rightmost panel shows exoplanet radius versus bulk density limited to planets with $\sigma_{\rho} < 0.1\rho$.}
    \label{fig:young}
\end{figure*}

Gl 436b, GJ 3470b, and HAT-P-11b all belong to field age systems with ages $>1$ Gyr. It has been suggested, based on recent constraints on the orbital structures of Gl 436b and GJ 3470b, that these older planets experiencing mass loss may have undergone late migration after some gravitational disturbance to its original far-away orbit \citep{2022A&A...663A.160B,2022ApJ...931L..15S}. HD 63433c is the only young planet with a detection of escaping neutral hydrogen. Two other young Neptunes, HD 63433b and K2-25b, have \lya\ transit observations, neither of which exhibit escaping neutral hydrogen \citep{2022AJ....163...68Z,2021AJ....162..116R}. This provokes interesting questions about the detectability or presence of atmospheric escape in young Neptune-sized planets. \cite{2021AJ....162..116R} describe multiple scenarios to reconcile the detected escaping atmospheres with the non-detections. For example, is the amount of ionizing radiation or strength of the stellar wind coming from a young star too great to allow a large planetary neutral hydrogen outflow to develop? These questions warrant more \lya\ transit observations of young exoplanet atmospheres.

\begin{deluxetable}{lccl}[t!]
\tablecaption{AU Mic system properties. \label{tab:prop}}
\tablecolumns{4}
\tablewidth{0pt}
\tablehead{
\colhead{Properties (Symbol)} &
\colhead{Value} &
\colhead{Units} &
\colhead{Reference}
}
\startdata
Earth-system distance (d) & $9.722100 \pm 0.0004625$ & pc & \cite{2018AA...616A...1G} \\
Age ($\tau$) & $22 \pm 3$ & Myr & \cite{2014MNRAS.445.2169M} \\
Right ascension ($\alpha$) & 20:45:09.87 & hh:mm:ss & ... \\
Declination ($\delta$) & $-$31:20:32.82 & dd:mm:ss & ... \\
Spectral type & M1Ve & & \cite{2015arXiv151001731T} \\
Bolometric luminosity (L$_{\text{bol}}$) & $0.09$ & L$_{\odot}$ & \cite{2009ApJ...698.1068P}\\
Stellar mass (M$_{\star}$) & $0.50 \pm0.03$ & M$_{\odot}$ & \cite{2020Natur.582..497P} \\
Stellar radius (R$_{\star}$) & $0.75 \pm0.03$ & R$_{\odot}$ & \cite{2019AAS...23325941W} \\
Stellar rotation period (P$_{\star}$) & $4.863 \pm0.010$ & days & \cite{1972ApL....11...13T} \\
Radial velocity (v$_{\star}$) & $8.7 \pm0.2$ & km s$^{-1}$ & \cite{2017AA...603A..54L} \\
Epoch (t$_0$) & $2458330.39080^{+ 0.00058}_{- 0.00057}$ & BJD & \cite{2022AJ....163..147G} \\
Transit duration & $3.56^{+0.60}_{-0.46}$ & hours & \cite{2022AJ....163..147G} \\
Planetary mass measurements (M$_{\text{p}}$) & $11.7^{+5.0}_{-5.0}$ & \Mearth & \cite{2022MNRAS.512.3060Z}$^{\text{a}}$ \\
... & $17.0^{+5}_{-5}$ & \Mearth & \cite{2021AA...649A.177M}$^{\text{b}}$ \\
... & $17.1^{+4.7}_{-4.5}$ & \Mearth & \cite{Klein2021}$^{\text{a}}$ \\
... & $20.12^{+1.57}_{-1.72}$ & \Mearth & \cite{2021AJ....162..295C}$^{\text{a}}$ \\
Planetary radius (R$_{\text{p}}$) & $4.19^{+0.24}_{-0.22}$ & \Rearth & \cite{2022AJ....163..147G} \\
Orbital period (P$_{\text{p}}$) & $ 8.4630004^{+0.0000058}_{-0.0000060}$ & days & \cite{2022AJ....163..147G} \\
Semi-major axis (a) & $0.0644^{+0.0056}_{-0.0054}$ & AU & \cite{2022AJ....163..147G} \\
\enddata
\tablenotetext{a}{Mass measured using radial velocity monitoring.}
\tablenotetext{b}{Mass measured using transit-timing variations.}
\end{deluxetable}

AU Mic is an M1Ve star in the $\beta$ Pic moving group \citep{2004ARA&A..42..685Z} with a reliably constrained age of $23 \pm 3$ Myr \citep{2014MNRAS.445.2169M}. At $9.79 \pm 0.04$ pc, it is one of the closest pre-main sequence stars to Earth. Two exoplanets transiting AU Mic have been discovered using data from {\it TESS} \citep{2022AJ....163..147G,2021AA...649A.177M,2020Natur.582..497P}. AU Mic b orbits closer to its host, with a period of $8.46$ days. An overview of AU Mic's system parameters, excluding AU Mic c, is given in Table~\ref{tab:prop}. \cite{2022AJ....163..147G} measured AU Mic b's radius as $4.19$ \Rearth, placing it on the edge of the hot Neptune desert as shown in the left-most panel of Figure~\ref{fig:young}. We expect planets of this size (indicating a large gaseous envelope) and orbital period (highly irradiated) to be experiencing significant photoevaporation. This and the star's youth and proximity to Earth prompt the detailed study of AU Mic b's potentially escaping atmosphere. We obtained \lya\ transits of AU Mic b with {\it HST}/STIS (HST-GO-15836; PI: Newton). These observations are presented in Section~\ref{sec:obs}. We analyze the presence and impact of an observed stellar flare in Section~\ref{sec:flare}. After accounting for the flare, we search for the presence of planetary absorption in \lya\ and other emission line light curves in Section~\ref{sec:lcurve}. We confirmed that AU Mic c does not transit during our observations. We also shifted the white-light mid-transit time for each visit (the zero hour mark in Figures~\ref{fig:flare},~\ref{fig:flmodel},~\ref{fig:lya_lc}, and~\ref{fig:metals_lc}) in line with recent transit-timing variation analyses for the AU Mic system \citep{2022AJ....164...27W}. We construct the high energy environment of AU Mic b, and estimate its mass loss rate and the impact of photoionization in Section~\ref{sec:highen}. We summarize and discuss our results in Section~\ref{sec:discuss}.


\section{Far-UltraViolet Observations} \label{sec:obs}

\subsection{New observations corresponding to two transits of AU Mic b}

We obtained FUV observations of AU Mic using a $0.2$\arcsec~x $0.2$\arcsec~aperture on {\it HST}'s STIS instrument. {\it HST} observed the AU Mic system on 2 July 2020 (Visit 1) and 19 October 2021 (Visit 2), corresponding to two transits of AU Mic b. Six science exposures were taken per transit, each exposure covering the observable window of an {\it HST} orbit ($2000-3000$ s). STIS was in its FUV multi-anode microchannel array configuration (FUV-MAMA). We used the E140M echelle grating, which has a resolving power of $\sim 45800$ and wavelength coverage from $1144-1729$ \AA. All science observations were made in TIME-TAG mode, where each photon detected by the instrument has a time-stamp. The time-stamps allowed us to split each science exposure into smaller sub-exposures prior to data extraction and reduction. This was done using the {\tt inttag} function from the {\tt stistools} Python package (v1.3.0)\footnote{\href{https://github.com/spacetelescope/stistools}{https://github.com/spacetelescope/stistools}}. For our flare analysis in Section~\ref{sec:flare}, we used $30$ s sub-exposures. Shorter sub-exposures were too low signal. We used 10 sub-exposures ($\sim 241$ or $275$ s each) per orbit for our light curve analysis in Section~\ref{sec:lcurve}. We extracted and reduced these data with {\tt stistools.x1d.x1d}, which utilizes the {\tt calstis} pipeline (v3.4). We allowed the pipeline to find the best extraction location per echelle order in the calibrated, flat-fielded exposures. The extraction locations per echelle order varied by less than a pixel between each orbit. The same extraction and reduction procedure was used on the sub-exposures. Each reduced spectrum contains 44 echelle orders.

The pipeline uncertainties correlate with $\sqrt{\text{N}}$, where N is the total photon count per wavelength bin. This uncertainty estimate is inaccurate for small counts, which is the case for the lower signal-to-noise sub-exposures. As in \cite{2021AJ....162..116R}, we recalculated the errors for each sub-exposure spectrum using Equation~\ref{eq:error}, an estimation of the confidence limit of a Poisson distribution, and the total counts reported in the {\tt GROSS} spectrum. These uncertainties were converted from counts into flux units using a multiplicative factor inferred by comparing the {\tt GROSS} and flux-calibrated spectra. The {\tt calstis} errors were used for the full-orbit spectra, which are high signal-to-noise.
\begin{equation}
    \sigma \approx 1 - \sqrt{\text{N}+0.75} \label{eq:error}
\end{equation}

\subsection{Investigating breathing}

\begin{figure*}
    \centering
    \includegraphics[width=\textwidth]{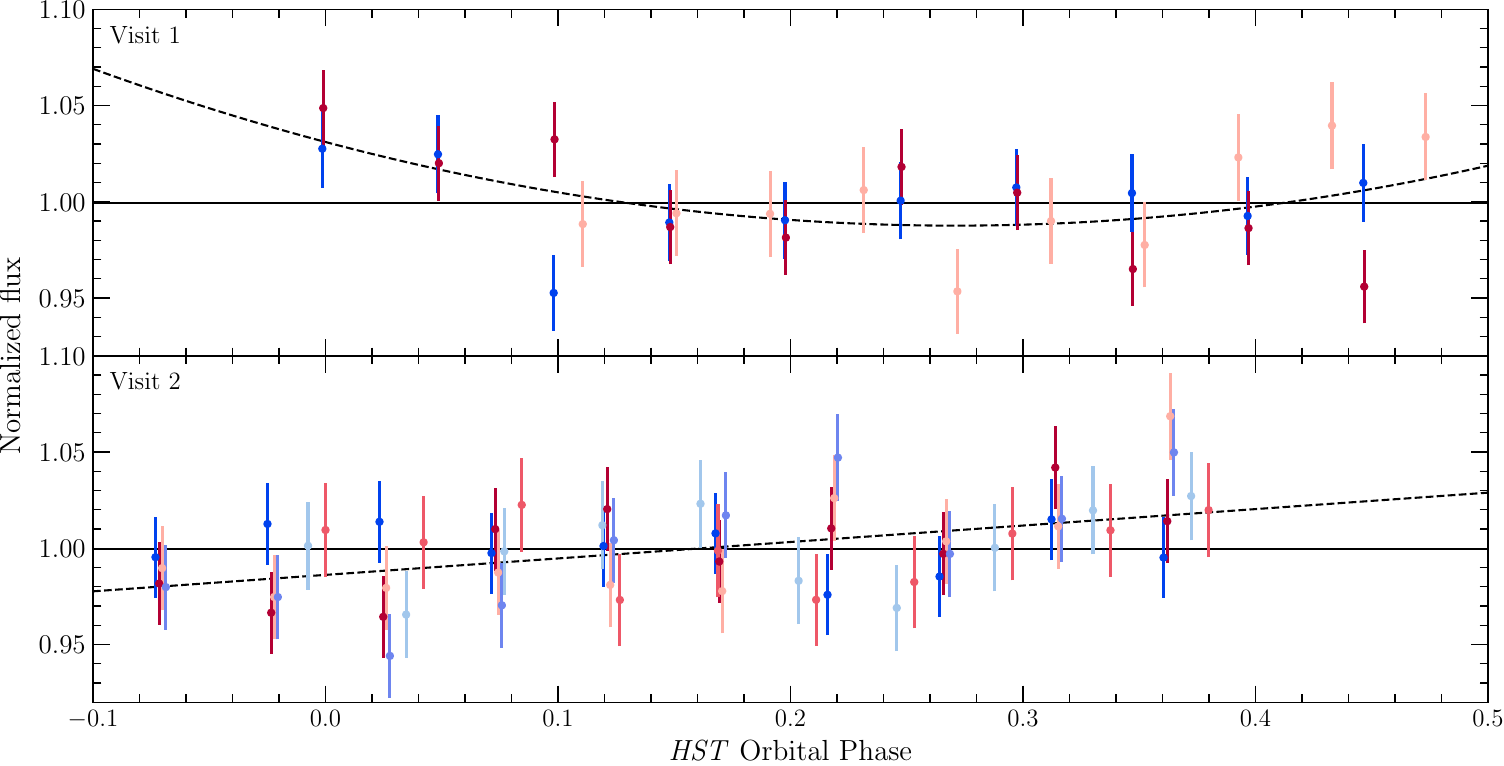}
    \caption{Visits 1 and 2 \lya\ light curves folded on the {\it HST} orbital period ($95.42$ min). Each color corresponds to sub-exposures from a single {\it HST} orbit. Only sub-exposures from consecutive orbits are plotted (and considered in fitting the trend). The best-fit polynomial is shown by the dashed line.}
    \label{fig:lya_folded}
\end{figure*}

Before analysis, we sought to remove {\it HST} ``breathing" - thermal fluctuations that cause variability in the telescope optics, impacting the throughput on the timescale of an {\it HST} orbit ($95.42$ min). For each transit, we created a light curve of consecutive orbits for the integrated \lya\ line flux in each sub-exposure ($10$ per orbit). \lya\ was integrated over $1214.0 - 1217.0$ \AA\ (excluding the remnant airglow region between $1215.4$ and $1215.7$ \AA). \lya\ is the highest signal-to-noise emission line in our data and thus the most preferred region to search for systematics. However, we note that our detrending process may overcorrect potential planetary and stellar variability on the -- relatively short -- timescale of an {\it HST} orbit.

Figure~\ref{fig:lya_folded} shows the \lya\ light curves for Visits 1 and 2 folded on {\it HST}'s period and the corresponding best-fit polynomials. The flux values are normalized by the mean flux per {\it HST} orbit. The best-fit per light curve was determined by selecting the polynomial order that gave the smallest Bayesian Information Criteria (BIC). Both transits deviate from a flat line, which corresponds to variability over the course of an {\it HST} orbit and confirms the presence of ``breathing". To detrend, we divided the best-fit polynomial out of each folded light curve (non-consecutive orbits included). The same process was used to detrend each FUV emission line light curve because {\it HST} breathing has been characterized as an achromatic effect. Most emission line light curves were consistent with a flat line when folded.

\subsection{Archival FUV observations}

We reference archival {\it HST}/STIS observations of AU Mic published by \cite{2000ApJ...532..497P}. Four science exposures were taken on 06 September 1998, which does not correspond to an AU Mic b transit (HST-GO-7556; PI: Linsky). The observing set-up was the same as our observations: $0.2$\arcsec~x $0.2$\arcsec~aperture, FUV-MAMA configuration, E140M grating, TIME-TAG mode. The same extraction, reduction, and breathing-check processes were used on these data.

\subsection{Looking for planetary absorption with \lya\ and other emission line spectra} \label{sec:spec}

\begin{figure*}
    \centering
    \includegraphics[width=\textwidth]{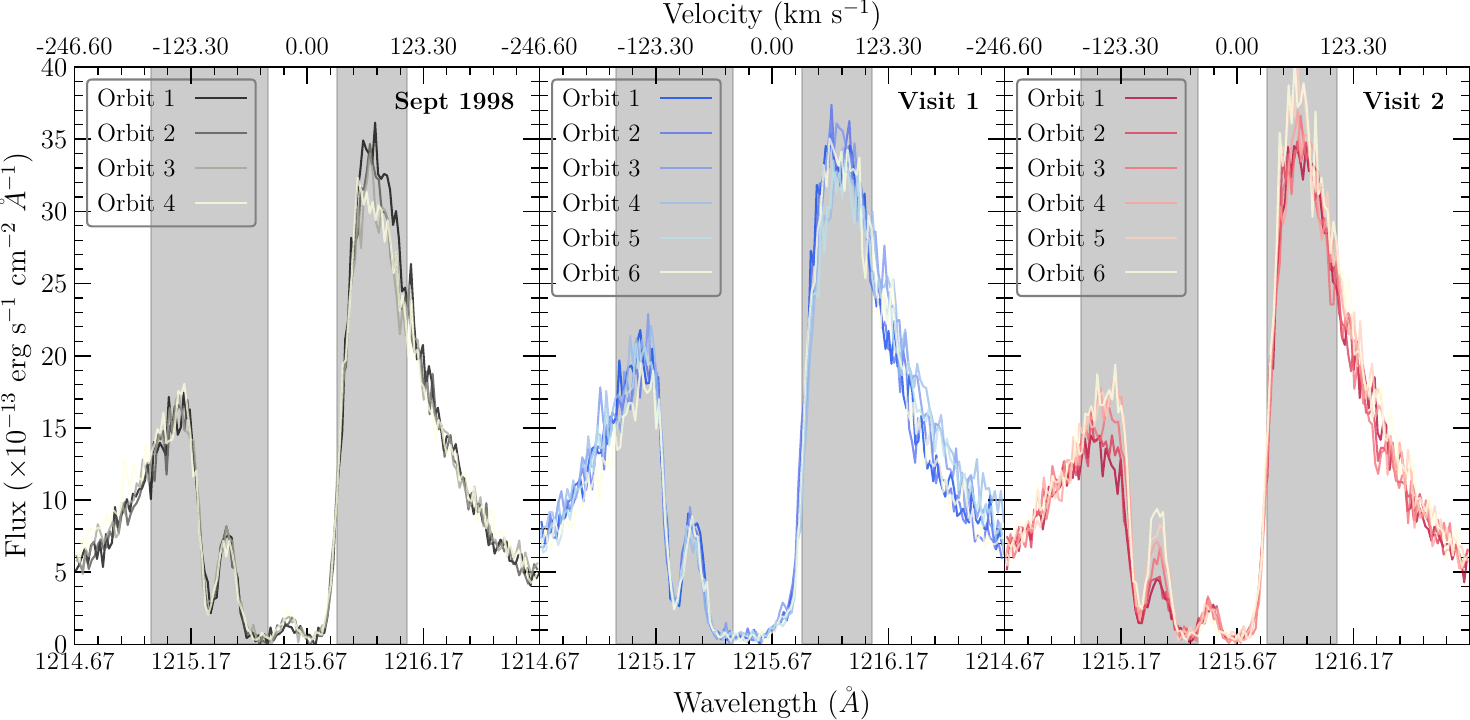}
    \caption{The \lya\ spectra of AU Mic observed by {\it HST}/STIS in September 1998 (left, archival), July 2020 (center, Visit 1), and October 2021 (right, Visit 2). The gray shaded areas indicate our blue- and red-wing integration regions.}
    \label{fig:lyaspec}
\end{figure*}

Figure~\ref{fig:lyaspec} depicts the observed \lya\ spectra for each new visit alongside the archival data. The ISM absorbs the line core ($\sim 1215.67$ \AA), leaving the wings for analysis. Although \lya\ observations of high radial velocity stars are becoming more common \citep{2022ApJ...926..129Y,2019ApJ...886...19S}, young stars in particular are low in velocity and as such do not experience significant Doppler shifting of spectral lines. AU Mic's low velocity ensures that the \lya\ line core is unobservable at Earth due to absorption by the ISM. Both the archival and Visit 2 data have residual geocoronal emission within the wavelengths fully absorbed by the ISM. Due to the high spectral resolution of STIS echelle data, this is not important for our analysis. 

Although the Visit 1 \lya\ spectra remain fairly consistent throughout the transit, the flare addressed in Section~\ref{sec:flare} is encapsulated in its first orbit. Due to the flare's short duration, it is washed out by the full-orbit exposure time and a flux increase is not visible. The flare still has the ability to impact our search for planetary absorption, however, and this first orbit cannot be considered devoid of any contamination despite its stable appearance. When discussing the full-orbit spectra, we ignore this orbit.

Over all three visits to AU Mic, the extended blue-wing ($\lambda < 1215$ \AA) remains stable. This may indicate that the blue-wing is relatively impervious to stellar activity. The same cannot be said for the extended red-wing ($\lambda > 1216.1$ \AA), which shows variations of up to $10\%$ percent in the archival, Visit 1, and Visit 2 data (see Figure~\ref{fig:lya_lc}). The inner red-wing -- where we are looking for planetary absorption ($1215.8 < \lambda < 1216.1$ \AA; $32.1 < v < 106$ km s$^{-1}$) -- shows similar fluctuations across the three visits. It may be that the entirety of the \lya\ red-wing is more sensitive to stellar activity than its blue counterpart. For that reason, we hesitate to associate a decrease in the \lya\ red-wing flux up to about $10\%$ with planetary absorption.

\begin{deluxetable}{lrr}
\tablecaption{Best-fit parameters with $1-\sigma$ uncertainties for the in-transit (Visit 2, first orbit) and out-of-transit (Visit 2, sixth orbit) \lya\ spectra: characterizing the intrinsic \lya\ line, the ISM absorption, and the planetary absorption. \label{tab:lyapy}}
\tablecolumns{3}
\tablewidth{0pt}
\tablehead{
\colhead{Parameter (Units)} &
\colhead{In-Transit Value} &
\colhead{Out-of-Transit Value}
}
\startdata
v$_{\text{n}}$ (km s$^{-1}$) & ... & $-7.29^{+0.92}_{-0.89}$ \\
log$_{10}$A$_{\text{n}}$ (erg s$^{-1}$ cm$^{-2}$ \AA$^{-1}$) & ... & $-10.87^{+0.03}_{-0.03}$ \\
FW$_{\text{n}}$ (km s$^{-1}$) & ... & $140.54^{+2.82}_{-2.74}$ \\
v$_{\text{b}}$ (km s$^{-1}$) & ... & $-7.37^{+1.18}_{-1.18}$ \\
log$_{10}$A$_{\text{b}}$ (erg s$^{-1}$ cm$^{-2}$ \AA$^{-1}$) & ... & $-11.64^{+0.02}_{-0.02}$ \\
FW$_{\text{b}}$ (km s$^{-1}$ ) & ... & $386.12^{+6.03}_{-5.87}$ \\
ISM log$_{10}$N$_{\text{H I}}$ & ... & $18.38^{+0.01}_{-0.01}$ \\
b (km s$^{-1}$) & ... & $12.34^{+0.27}_{-0.29}$ \\
v$_{\text{H I}}$ (km s$^{-1}$) & ... & $-22.98^{+0.23}_{-0.23}$ \\
pl log$_{10}$N$_{\text{H I}}$ & $13.96^{+0.03}_{-0.03}$ & ... \\
b$_{\text{pl}}$ (km s$^{-1}$) & $61.05^{+2.24}_{-2.24}$ & ... \\
v$_{\text{pl}}$ (km s$^{-1}$) & $-61.26^{+2.45}_{-2.74}$ & ... \\
\enddata
\end{deluxetable}

We expect the majority of escaping planetary material to be moving at $10$s of km s$^{-1}$. Hydrodynamic models show that radiation and stellar wind pressure can accelerate escaping neutral hydrogen up to these velocities directed radially outward from the host star \citep{2019ApJ...873...89M,2018A&A...620A.147B}. This draws our attention to the inner blue-wing ($1215 < \lambda < 1215.5$ \AA; $-165.2 < v < -41.9$ km s$^{-1}$) which could be experiencing flux variations due to absorption by planetary neutral hydrogen flowing away from the host star. The archival spectra do not show any significant changes within this region. The Visit 1 spectra display a small amount of variability at the shorter wavelength edge of our blue-wing region. The Visit 2 spectra, however, show a gradual $30\%$ increase in the blue-wing flux over the course of the planet's transit.

Assuming the flux behavior is due to absorption by escaping planetary neutral hydrogen, we fit the observed \lya\ profile with an intrinsic stellar component, an ISM absorption component, and a planetary absorption component. We used the Python package {\tt lyapy} \citep{2022zndo...6949067Y}, which encompasses a model that combines the aforementioned emission and absorption components and a fitting algorithm that uses {\tt emcee} as the MCMC sampler \citep{2013PASP..125..306F}. A thorough description of the {\tt lyapy} model and how it can be used to reconstruct cool star \lya\ emission profiles can be found in \cite{2016ApJ...824..101Y} and Section 4.1 of \cite{2020AA...637A..22R}. 

\begin{figure*}[t!]
    \centering
    \includegraphics[width=\textwidth]{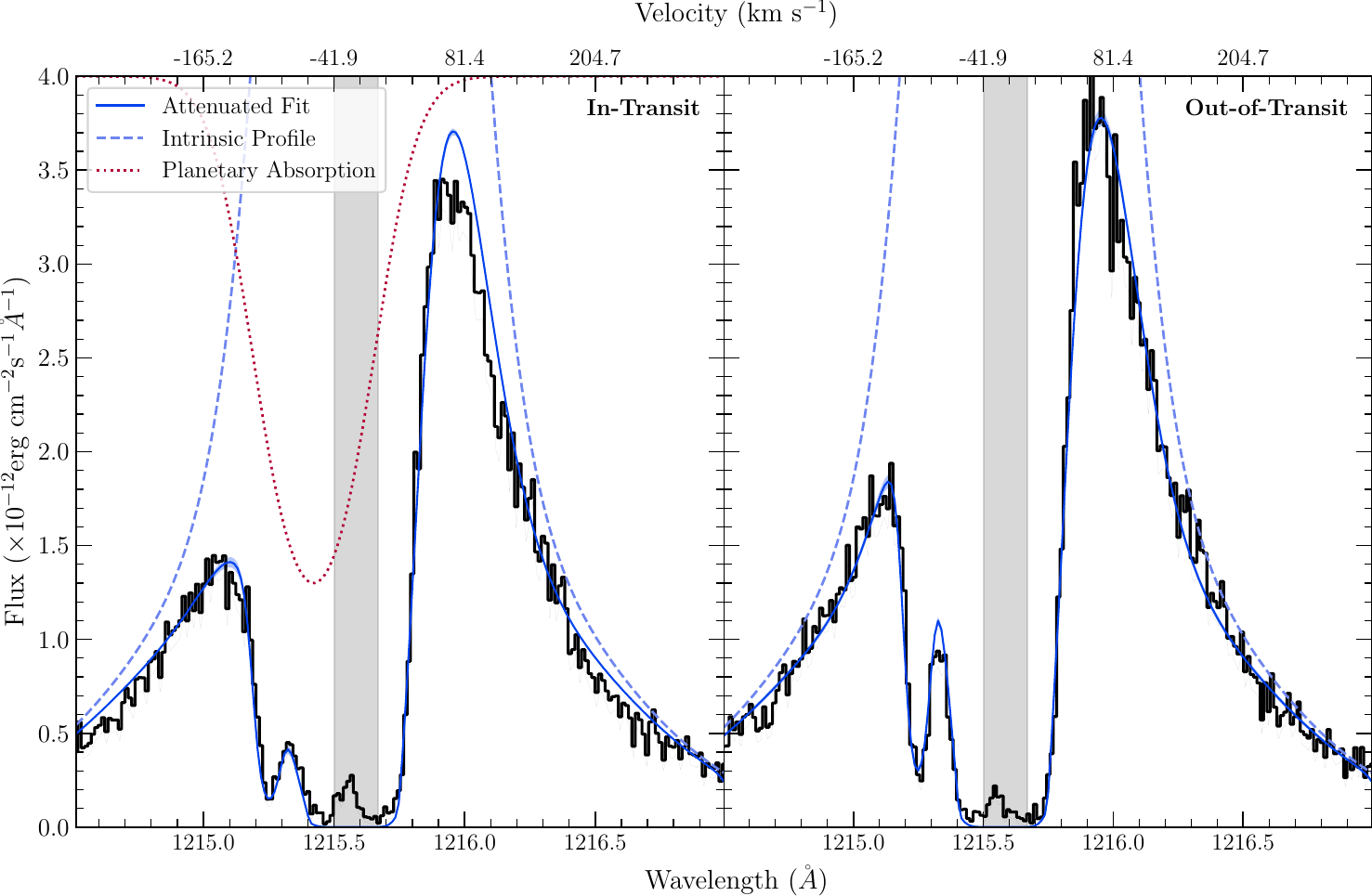}
    \caption{The AU Mic \lya\ spectra ``in-transit" on the left (Visit 2 first orbit) and ``out-of-transit" on the right (Visit 2 sixth orbit). The observed data is shown by the black line. The blue dashed line is the reconstructed stellar \lya\ line. The Voigt profile fit for the planetary absorption scaled to the y-axis of the figure is represented by the red dotted line. The blue line is the best fit damped \lya\ profile, with its $1-\sigma$ uncertainty shown by shaded blue regions around the line. The gray shaded regions indicate areas of the data that were masked before being run through our fitting algorithm.}
    \label{fig:lyapyspec}
\end{figure*}

First, we reconstructed AU Mic's \lya\ emission line using the last orbit from Visit 2 (see right panel of Figure~\ref{fig:lyapyspec}), which we assume is out-of-transit for the planetary neutral hydrogen. The timing of absorption in comparison to the planet's white-light transit is discussed more thoroughly in Sections~\ref{sec:lcurve} and~\ref{sec:discuss}. We varied every stellar and ISM profile parameter with uniform priors except for the ISM deuterium-to-hydrogen ratio, which we fixed to $1.5\times 10^{-5}$ \citep{2006ApJ...647.1106L}. The best-fit parameters that recreated our observed \lya\ profile can be found in Table~\ref{tab:lyapy}. Our results do not agree within uncertainties with the reconstructed AU Mic \lya\ profile from \cite{2016ApJ...824..101Y} (see their Table 1). However, our best-fit values agree within a few $\sigma$ at most and the stability of M dwarf \lya\ emission at young ages has not been fully explored.

We modified the absorption model within {\tt lyapy} to include a Voigt profile characterizing the excess absorption due to the proposed planetary outflow. {\it This profile is only able to characterize the high velocity outflow that is visible in our observations.} All information regarding the low velocity planetary material is lost due to ISM absorption. The profile is parameterized by column density, Doppler broadening, and a velocity centroid associated with the intervening planetary neutral hydrogen. These three parameters were varied during our fit to the in-transit spectrum, whereas our best-fit \lya\ emission and ISM absorption values from the out-of-transit profile were fixed. We estimate the high velocity planetary neutral hydrogen has a column density of log$_{10}$N$_{\text{H I}} = 13.96^{+0.03}_{-0.03}$ and is moving $-61.26^{+2.45}_{-2.74}$ in the stellar rest frame. Our column density should be treated as a lower limit on the total neutral hydrogen column density of the outflow. The left panel of Figure~\ref{fig:lyapyspec} shows the best-fit planetary absorption scaled to the figure's y-axis along with the best-fit to the observed spectrum. We investigate this in-transit absorption further by integrating over the region and analyzing the light curve in Section~\ref{sec:lcurve}.

\begin{figure*}
    \centering
    \includegraphics[width=\textwidth]{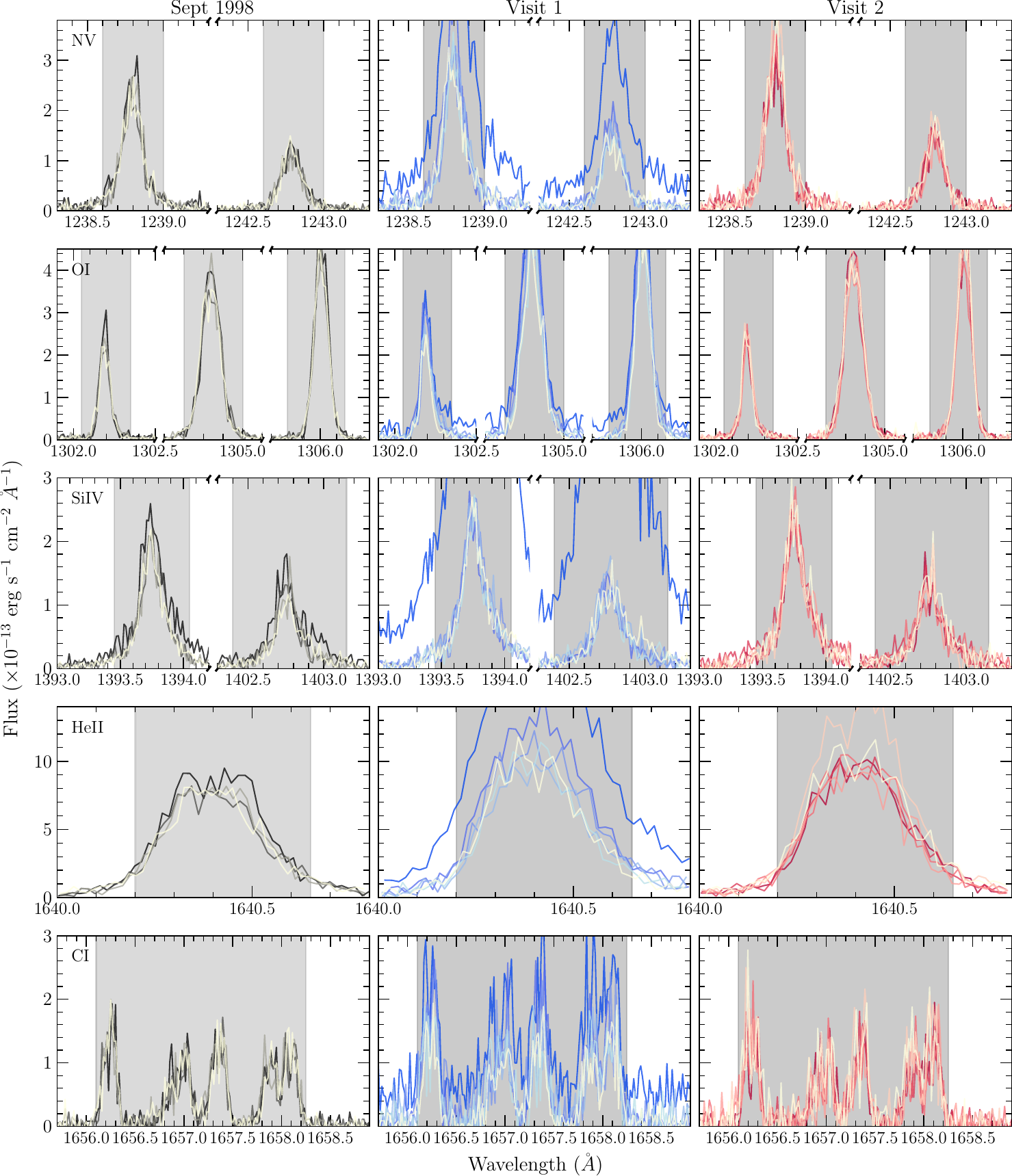}
    \caption{The emission line spectra of AU Mic observed by {\it HST}/STIS in September 1998 (left, archival), July 2020 (center, Visit 1), and October 2021 (right, Visit 2). The gray shaded areas indicate our integration regions. The orbits correspond to the same colors indicated in Figure~\ref{fig:lyaspec}.}
    \label{fig:metalspec}
\end{figure*}

Figure~\ref{fig:metalspec} shows the archival, Visit 1, and Visit 2 spectra for the \ion{N}{5}, \ion{O}{1}, \ion{Si}{4}, \ion{He}{2}, and \ion{C}{1} species. Although there are more emission lines present in the data (see Table~\ref{tab:lines} and Figures~\ref{fig:full_spec1} and~\ref{fig:full_spec2}), we focus on these five species for tracing stellar activity and our search for planetary atmospheric escape. FUV spectroscopic studies of M dwarfs have shown that \ion{N}{5} and \ion{Si}{4} can trace stellar activity, which can help identify when flux changes are not planetary in nature \citep{2018ApJ...867...71L,2014ApJS..211....9L}. Alternatively, \ion{O}{1}, \ion{He}{2}, and \ion{C}{1} are relatively insensitive to flares. \ion{O}{1} and \ion{C}{1} are of recent interest as potential tracers of the evolution of the C/O ratio in exoplanet atmospheres \citep{2022BAAS...54e2805O}. \ion{O}{1}, \ion{C}{1} and \ion{C}{2} have been used to trace atmospheric escape previously in several hot Jupiters, as well as the hot super Earth $\pi$ Men c \citep{2021ApJ...907L..36G}.

As expected, the Visit 1 flare causes substantial increases in the emission line fluxes for \ion{N}{5} and \ion{Si}{4}, whereas \ion{O}{1} and \ion{C}{1} do not. \ion{He}{2} also experiences an enhancement due to the flare. In Visit 2, \ion{N}{5} and \ion{He}{2} have a similar flux increase during the fifth orbit, which also corresponds to the increase observed in the \lya\ blue-wing in Figure~\ref{fig:lyaspec} but does not last through the sixth orbit. When compared to the archival spectra, this behavior is not representative of the typical noise in \ion{N}{5} and \ion{He}{2} emission. Light curves for these two species are presented in Section~\ref{sec:lcurve} and compared to the \lya\ line behavior. \ion{O}{1} shows consistent spectral variation across the archival, Visit 1, and Visit 2 spectra, making it difficult to discern the presence of planetary absorption. Similarly, \ion{C}{1} is too low signal-to-noise to draw any conclusion about carbon in transmission. Their light curves are also shared in Section~\ref{sec:lcurve}.

\begin{deluxetable}{cllr}[b!]
\tablecaption{AU Mic FUV emission lines confidently identifiable in this work's {\it HST}/STIS spectra, cross-referenced with NIST \citep{NIST_ASD}, the CHIANTI Atomic Database \citep{2021ApJ...909...38D,1997AAS..125..149D}, and \cite{2022AJ....164..110F}. The table includes line location(s), integration region, and formation temperature. \label{tab:lines}}
\tablecolumns{4}
\tablewidth{0pt}
\tablehead{
\colhead{Species} &
\colhead{$\lambda$ (\AA)} &
\colhead{$\Delta\lambda$ (\AA)} &
\colhead{$\log_{10}(T/\text{K})$}
}
\startdata
\ion{C}{3} & $1175.7$ & $1174.5 - 1177.0$ & $4.8$ \\
\ion{Si}{3} & $1206.51$ & $1206.3 - 1206.7$ & $4.7$ \\
\ion{H}{1} (\lya) & $1215.67$ & $1214.0 - 1217.0$ & $4.5$ \\
\ion{N}{5} & $1238.82$, $1242.80$ & $1238.6 - 1239.0$, $1242.6 - 1243.0$ & $5.2$ \\
\ion{O}{1} & $1302.2$, $1304.84$, $1306.01$ & $1302.05 - 1302.35$, $1304.7 - 1305.0$, $1305.8 - 1306.15$ & $3.8$ \\
\ion{C}{2} & $1334.53$, $1335.71$ & $1334.45 - 1334.75$, $1335.45 - 1335.95$ & $4.5$ \\
\ion{Fe}{21} & $1354.07$ & $1353.8 - 1354.4$ & $7.1$ \\
\ion{Si}{4} & $1393.76$, $1402.77$ & $1393.45 - 1394.05$, $1402.4 - 1403.15$ & $4.9$ \\
\ion{C}{4} & $1548.20$, $1550.774$ & $1547.9 - 1548.5$, $1550.6 - 1551.0$ & $4.8$ \\
\ion{He}{2} & $1640.40$ & $1640.2 - 1640.65$ & $4.9$ \\
\ion{C}{1} & \begin{tabular}{l}$1656.27$, $1656.93$, $1657.01$, $1657.38$, \\$1657.91$, $1658.12$\end{tabular} & $1656.0 - 1658.25$ & $3.8$ \\
\enddata
\end{deluxetable}

\section{Flare Analysis} \label{sec:flare}

\subsection{Emission line flare light curves}

\begin{figure*}[t!]
    \centering
    \includegraphics[width=\textwidth]{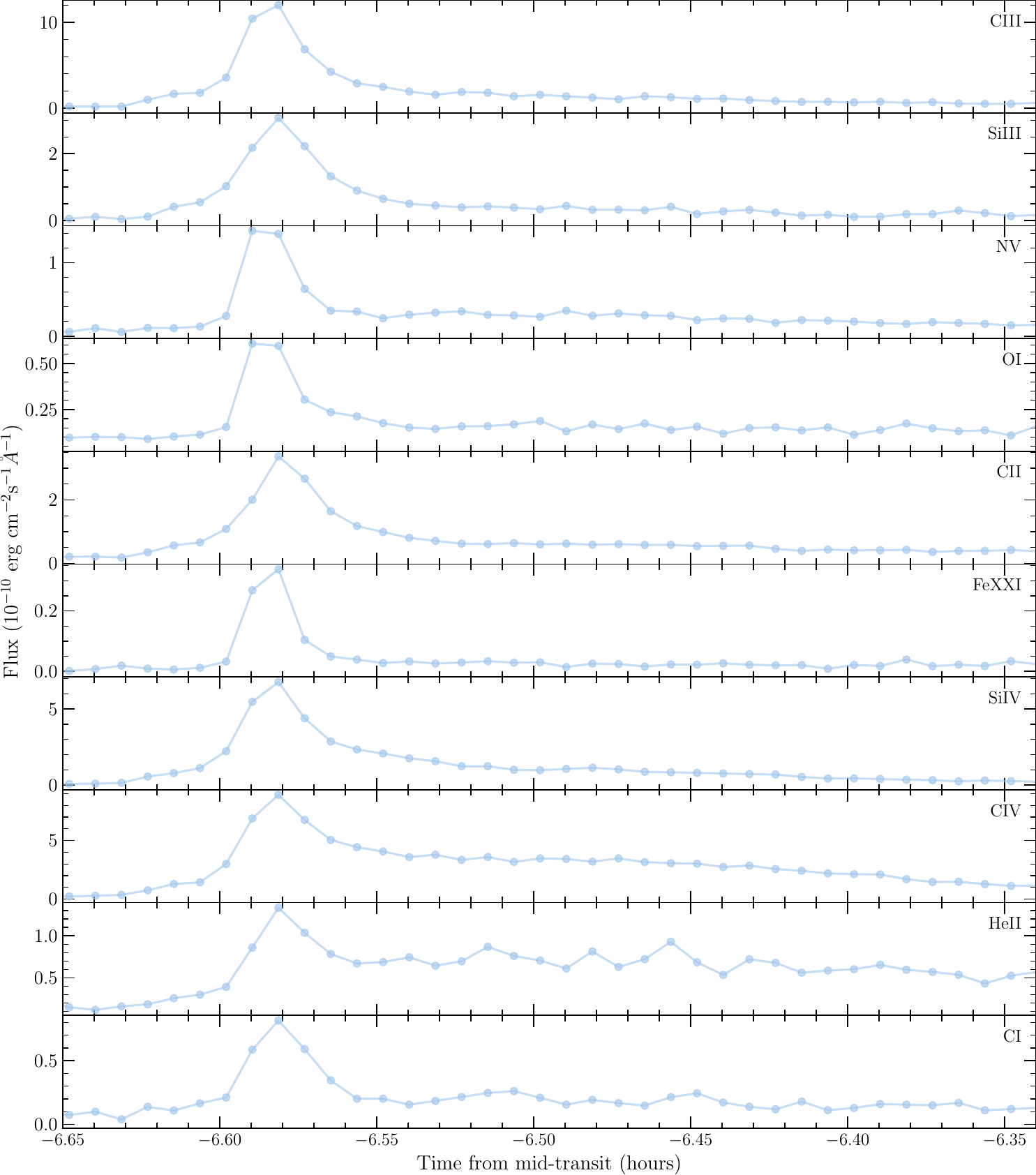}
    \caption{Light curves for the first orbit of Visit 1 split into $30$ s sub-exposures for various emission lines from Table~\ref{tab:lines}.}
    \label{fig:flare}
\end{figure*}

Before looking for a planetary signature, we needed to characterize and remove any host star variability from our observations. A substantial increase in AU Mic's flux in several emission lines and its FUV continuum occurred during the first orbit of Visit 1, initially identified by eye. Figure~\ref{fig:flare} shows the diversity in flaring behavior of the highest signal-to-noise emission lines (listed in Table~\ref{tab:lines}). We used $30$ s sub-exposures for comprehensive sampling of the flare. We did not model these light curves as we did with \lya\ and the continuum presented in Section~\ref{sec:lyaflare}.

As shown in Figure~\ref{fig:flare}, \ion{C}{3}, \ion{Si}{4} and \ion{C}{4} exhibit the strongest emission during the peak of the flare, followed by \ion{Si}{3} and \ion{C}{2}. Most emission lines return to quiescent fluxes relatively quickly ($\sim 2$ min). A notable exception is \ion{He}{2}, which remains elevated over the entirety of the exposure and returns to quiescence by the second exposure (not pictured). The Visit 1 flare caused greater flux increases in every overlapping emission line examined between this work and \cite{2022AJ....164..110F}. However, caution should be taken when comparing fluxes observed with STIS and COS, as STIS has focus issues that impact the flux calibration of narrow slit observations \citep{2018stis.rept....6R,2017stis.rept....1P}. The one flare we observe in our $\sim 7$ hours of non-consecutive {\it HST} orbits is at odds with the $2.5$ hour$^{-1}$ AU Mic FUV flare rate presented in \cite{2022AJ....164..110F} and potentially agrees with the $2$ day$^{-1}$ estimate from \cite{2022AJ....163..147G} based on visible light data. Although we did not conduct further analysis, characterizing this flare and comparing it to other AU Mic flares observed with STIS and COS would further knowledge of flare diversity from an individual star and help identify systematic differences between the STIS and COS instruments.

\subsection{\lya\ and continuum flare modeling} \label{sec:lyaflare}

Flares and other forms of stellar activity can mask or masquerade as planetary signals in transit photometry and transmission spectroscopy. \cite{2014ApJS..211....9L} surveyed 38 cool stars with {\it HST}/COS and STIS spectra, characterizing their variability in the ultraviolet and the potential for masking planetary signals. They concluded that most F-M stars will allow for confident detections of planets larger than Jupiter, with M dwarfs potentially lowering that limit to Neptune-to-Saturn-sized planets.

We modeled the Visit 1 flare in order to characterize its potential to mask the transit of AU Mic b. For this analysis, we focused on the \lya\ emission line because it is the highest signal emission line in the data and because of its relevance to our search for atmospheric escape in transmission. We also modeled the FUV continuum behavior during the flare for comparison (continuum defined in~\ref{ap:cont}).

\begin{figure*}[t!]
    \centering
    \includegraphics[width=\textwidth]{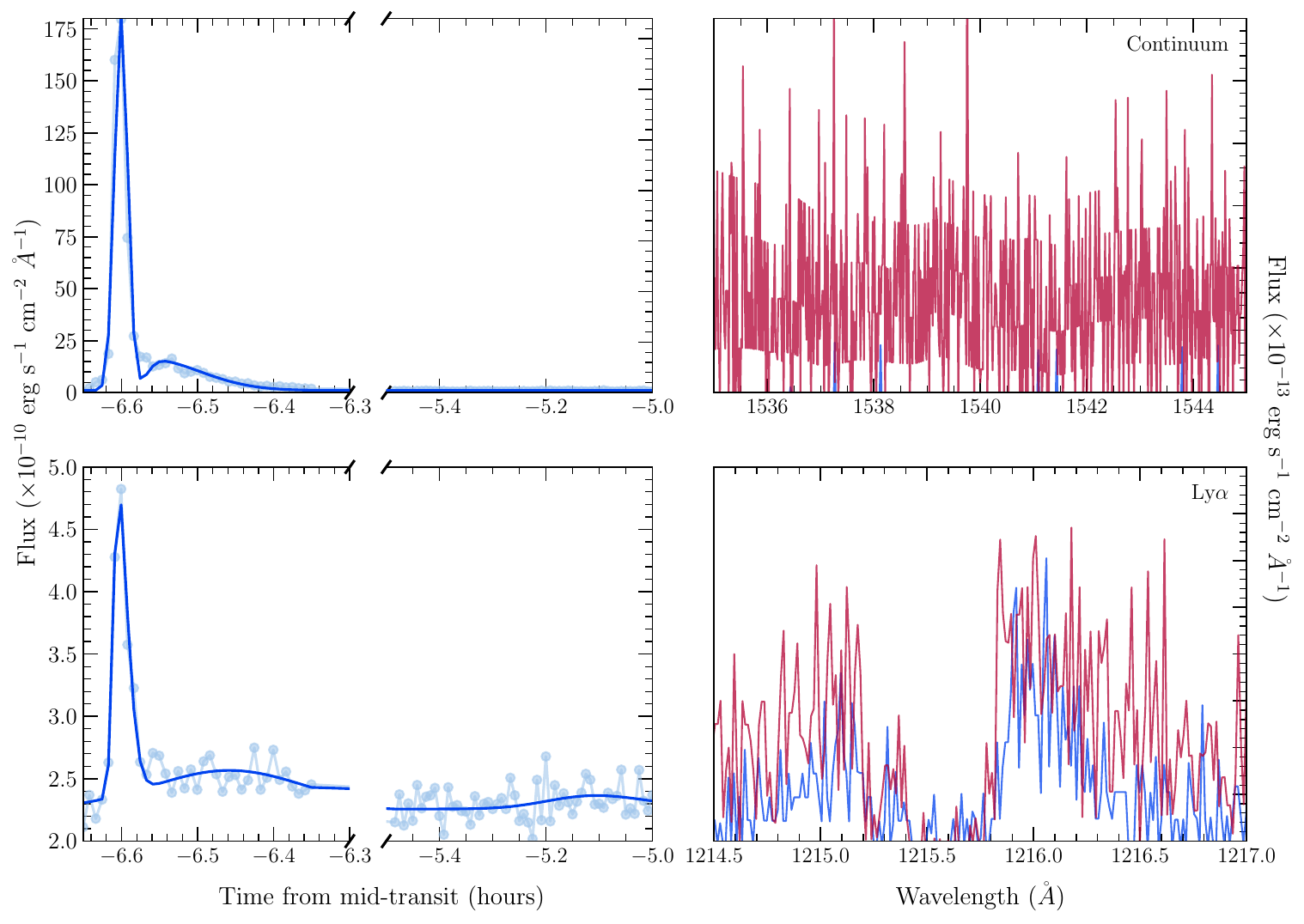}
    \caption{Continuum (top, left) and \lya\ (bottom, left) light curves for the first two orbits of Visit 1. The light blue points are the $30$ s sub-exposures, and the blue line is the fitted flare model. Continuum and \lya\ in- (red) and out-of-flare (blue) spectra are shown in the top-right and bottom-right panels, respectively.}
    \label{fig:flmodel}
\end{figure*}

We adapted the open source Python code {\tt cos\_flares} from \cite{2022AJ....164..110F}\footnote{\href{https://github.com/afeinstein20/cos_flares/tree/paper-version}{https://github.com/afeinstein20/cos\_flares/tree/paper-version}} written for analyzing flares in {\it HST}/COS data to be compatible with our STIS data. We fit the Visit 1 flare light curve using a skewed Gaussian, represented by Equation~\ref{eq:sgflare} where $\omega$ parameterizes the amplitude, $t$ is time, $\xi$ is the average time in the flare light curve, and $\eta$ is a dimensionless version of time related to the aforementioned parameters and the skewness of the light curve. The skewed Gaussian model was chosen because it provided the best fit to the \lya\ light curve compared to the white-light flare  model and the skewed Gaussian convolved with the white-light flare model. Figure~\ref{fig:flmodel} shows the resulting fit to the \lya\ flare. We allowed for multiple peaks in the flare to be modeled and fit, which is shown by the small increase in flux after the flare.

\begin{equation}
    F (t) \sim \frac{1}{\omega\sqrt{2\pi}}\left(e^{-(t - \xi)^2/(2\omega)^2}\right)\left[1+\text{erf}\left(\frac{\eta}{\sqrt{2}}\right)\right] \label{eq:sgflare}
\end{equation}

\subsection{Flare energy and duration}

We calculated the flare energy, equivalent duration, and maximum time the flare could mask a $10\%$ planetary transit. 

Absolute flare energy, $E$, is the total energy released by the flare at the star within the spectral feature. It is given by Equation~\ref{eq:flenergy}:
\begin{equation}
    E = 4\pi d^2 \int \left(F_{f}(t) - F_q(t)\right)dt \label{eq:flenergy}
\end{equation}
where $d$ is the distance to AU Mic, and $F_{fl}$ and $F_q$ are the flaring and quiescent fluxes \citep{2014ApJ...797..122D,2014ApJ...797..121H}. The difference in flaring and quiescent flux is integrated over the time it takes for the flux to return to quiescence. We classified the first two orbits of Visit 1 as flaring, and the remaining out-of-transit orbit at the end of the visit as quiescent.

We calculated a total flare energy of $7.11\times 10^{32}$ erg for \lya, and $2.57\times 10^{34}$ erg for the continuum. Our continuum flare energy is $100$ times larger than the median flare energy characterized in \cite{2022AJ....164..110F} for AU Mic. It is also $1000 - 10000$ times larger than the median FUV flare energy found by surveying M stars \citep{2018ApJ...867...70L,2018ApJ...867...71L}.

The equivalent duration, $t_{eq}$, is given by Equation~\ref{eq:fldur}. Similar to an equivalent width, it characterizes the amount of time for which the star would have to emit 100\% of its quiescent flux in order to match the total energy of the flare. The flare's equivalent duration was estimated as $278$ s ($0.0772$ hours) for \lya, and $43515$ s ($12.1$ hours) for the continuum.

\begin{equation}
    t_{eq} = \int \left(\frac{F_{f}(t) - F_q(t)}{F_q(t)}\right)dt \label{eq:fldur}
\end{equation}

The maximum masking time bears the most importance to our search for AU Mic b in transmission. We calculated the amount of time the flare was above $110\%$ of the quiescent \lya\ flux, effectively washing out a $10\%$ planetary transit depth. We chose $10\%$ as a conservative estimate for the potential size of the exosphere. We found that this flare contaminated the \lya\ light curve for about $5731$ s, or $1.5$ {\it HST} orbits. This occurred 7 hours before AU Mic b's white-light mid-transit, so we do not expect the increase in flux from the flare to deter us from observing a planetary signal in Visit 1.


\section{AU Mic's Light Curves} \label{sec:lcurve}

We examined the temporal behavior of AU Mic's flux in different regions of its FUV spectrum (i.e., the \lya\ emission line and the several metal lines listed in Table~\ref{tab:lines}). In Section~\ref{sec:flare} we concluded that the flare only impacts the first orbit of Visit 1, so we removed the contaminated orbit and examined the Visit 1 and 2 light curves separately for signs of planetary absorption.

\subsection{\lya\ light curve} \label{sec:lyalcurve}

\begin{figure*}[b!]
    \centering
    \includegraphics[width=\textwidth]{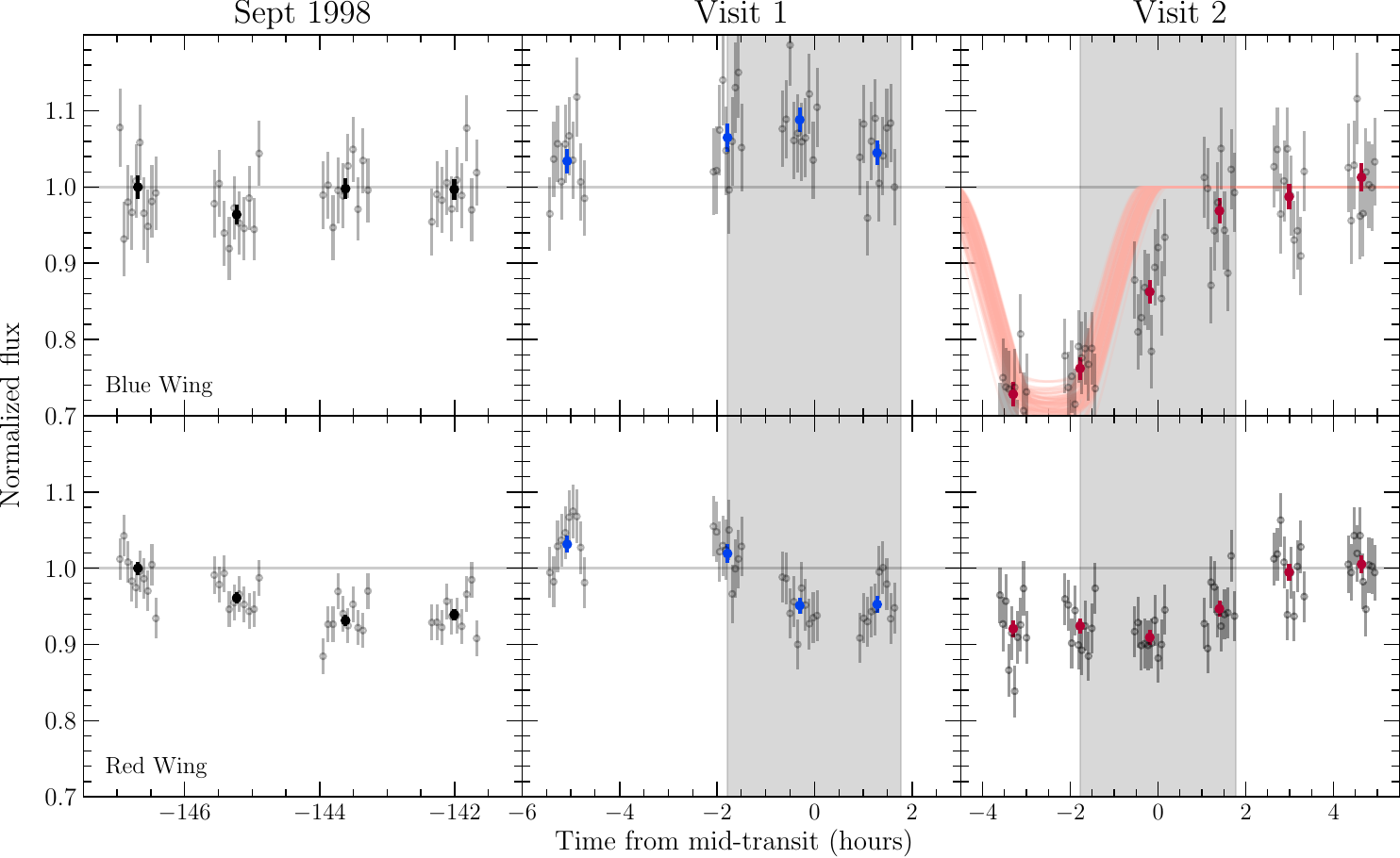}
    \caption{The archival (left, black), Visit 1 (center, blue), and Visit 2 (right, red) light curves for the blue- (top) and red-wing (bottom) of the \lya\ line. The first orbit from Visit 1 is not included because of the flare. Visit 1's sixth, and final, orbit is not included but is represented by the horizontal line at a flux of 1. The gray points represent the sub-exposures. Samples from the posterior distribution of light curve fits to the Visit 2 blue-wing are shown by the pink lines. The white-light transit duration is shown by the gray shaded region.}
    \label{fig:lya_lc}
\end{figure*}

The presence or absence of absorption in the \lya\ line over the course of AU Mic b's transit can indicate whether or not neutral hydrogen is escaping from the planet's upper atmosphere. After the detrending process described in Section~\ref{sec:obs}, we created two \lya\ light curves per transit by summing the flux in the blue-wing ($1215.0 - 1215.5$ \AA) and the red-wing ($1215.8 - 1216.1$ \AA). The errors were summed in quadrature. Visit 1 exposures were normalized by the sixth orbit, which occurs $17$ hours after mid-transit and $>20$ hours after the flare. Visit 2 was normalized by the average of the fifth and sixth orbits. We also included archival {\it HST}/STIS observations of AU Mic from September 1998 which do not coincide with an AU Mic b or c transit, normalized by the last orbit from this visit. These light curves are shown in Figure~\ref{fig:lya_lc}.

The \lya\ blue-wing shows a $\sim 5\%$ increase in emission during Visit 1 (center-top panel of Figure~\ref{fig:lya_lc}). For this reason, we do not attempt to fit for a planetary transit in this light curve. We present possible explanations for this in Sections~\ref{sec:metals} and~\ref{sec:discuss}. As discussed in Section~\ref{sec:obs}, the \lya\ red-wing exhibits consistent flux changes of up to $10\%$ across all three {\it HST} visits. These fluctuations occur at different times in relation to the planetary ephemeris -- from $100$s to within a few hours outside of transit -- leading us to conclude that these are changes in the intrinsic stellar \lya\ emission.

The Visit 2 \lya\ blue-wing light curve exhibits extended absorption starting at least four hours prior to the white-light mid-transit. The blue-wing flux decrease lasts for about four hours and is completely recovered by the conclusion of white-light transit. If the decrease in flux is due to absorption in AU Mic b's atmosphere, that implies that the neutral hydrogen is moving {\it ahead} of the planet and is expanding radially away from the host star. We pose physical explanations for the bulk of the escaping atmosphere to transit before the planet in Section~\ref{sec:discuss}. Although slight \lya\ absorption prior to planet transit has been observed in other systems, like GJ 3470b, it has never been observed to this magnitude \citep{2018A&A...620A.147B}.

\begin{deluxetable}{cll}
\tablecaption{Fixed Light Curve Parameters \citep{2022AJ....163..147G}.
\label{tab:batman}} 
\tablecolumns{3}
\tablewidth{0pt}
\tablehead{
\colhead{Parameter} &
\colhead{Value} &
\colhead{Units}
}
\startdata
$P_{\text{p}}$ & $8.4630004$ & days \\
$a$ & $0.0644$ & AU \\
$i$ & $89.18$ & degrees \\
$e$ & $0.12$ & \\
$w$ & $88.528$ & degrees \\
\enddata
\end{deluxetable}

We used {\tt batman} to model a planetary transit for the Visit 2 blue-wing and {\tt emcee} to explore parameter space using Markov Chain Monte Carlo (MCMC). {\tt batman} models a transit for a fully opaque circle, not allowing for more diverse atmospheric geometries. We initialized the {\tt batman} model with measured system parameters (see Table~\ref{tab:batman}). Only the planetary radius at \lya\ and the mid-transit time were free parameters. The mid-transit time had a uniform prior constrained to within 5 hours of the white-light mid-transit time. The planetary radius at \lya\ was limited to positive values (no negative transit depths). The 60 sub-exposures associated with Visit 2 were fit using 80 MCMC walkers and 200000 steps. There was a burn-in of 50000 steps. We did the same for the Visit 2 red-wing, but the opaque disk model provides a poor fit to our data due to its long duration and small flux decrease.

The best-fit $R_{\text{p}}/R_{\star}$ value for the blue-wing light curve is $0.52 \pm 0.01$ ($0.39 \pm 0.008$ \Rsol). The \lya\ blue-wing mid-transit time is $2.5 \pm 0.24$ hours before the white-light mid-transit time. Figure~\ref{fig:lya_lc} includes example light curves pulled from the MCMC posteriors as pink lines. A model that allows for non-circular geometries and/or transparency of the transiting object along with data corresponding to the beginning of the absorption signal would be required to better fit our blue-wing light curve.

We used the formalism presented in \cite{2023MNRAS.518.4357O} to estimate the geometry and kinematics of the escaping planetary material. Although they model a trailing tail of neutral hydrogen, the same physics applies to a leading tail -- our observations of AU Mic b correspond to a flipped version of Figure 1 from \cite{2023MNRAS.518.4357O}. We used their Equations 10 and 11 for transit depth and duration, respectively, to calculate a tail length of $9.69 \times 10^{10}$ cm ($1.39$ \Rsol) and height of $2.22 \times 10^{10}$ cm ($0.32$ \Rsol). From their Equation 2, we estimated the velocity of the bulk of escaping material -- before acceleration by interactions with the environment -- to be $20$ km s$^{-1}$. Unfortunately we are unable to probe this velocity observationally for the AU Mic system due to the ISM absorption at the \lya\ core (see Section~\ref{sec:spec}), but \lya\ observations of high radial velocity systems allow transits to occur at these lower velocities, as do transits of uncontaminated emission lines (e.g., He $10830$ \AA). As specified by \cite{2023MNRAS.518.4357O}, we treat these calculations as rough estimates to be tested by future sophisticated models.

\subsection{Metal line light curves} \label{sec:metals}

\begin{figure*}[t!]
    \centering
    \includegraphics[width=\textwidth]{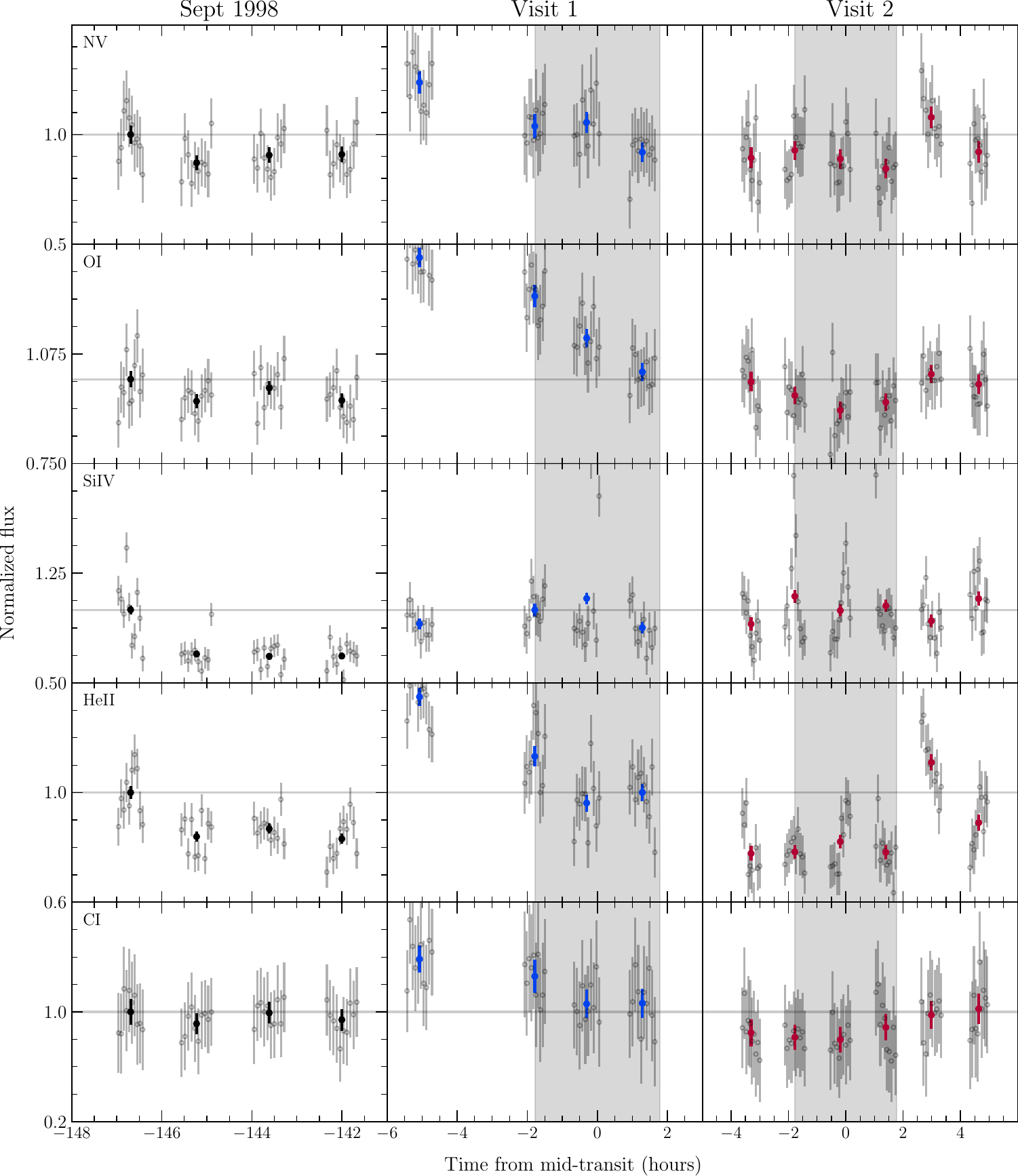}
    \caption{The archival (left, black), Visit 1 (center, blue), and Visit 2 (right, red) light curves for several FUV emission lines. The first orbit from Visit 1 is not included because of the flare. Visit 1's sixth, and final, orbit is not included but is represented by the horizontal line at a flux of 1. The gray points represent the sub-exposures. The white-light transit duration is shown by the gray shaded region.}
    \label{fig:metals_lc}
\end{figure*}

The search for atomic metals has been ongoing in some of the hottest giant planet atmospheres. Detections of atomic sodium, potassium, and calcium have come from optical and infrared observations \citep[e.g.,][]{2022A&A...666A..47Z,2020A&A...641A.123H,2019A&A...632A..69Y}. Using {\it HST}/STIS, neutral and ionized metals (mostly sodium) have been detected in the ultraviolet for several Hot Jupiters \citep{2019AJ....158...91S,2018AJ....156..298A,2016Natur.529...59S,2014MNRAS.437...46N}. With {\it HST}/COS, metal absorption extending past the Roche lobe has been detected for WASP-12b, showing that these ionized metals can be carried far away from the planet by atmospheric escape \citep{2012ApJ...760...79H,2010ApJ...714L.222F}. Hot Neptunes, despite being more vulnerable to an escaping atmospheric outflow that could carry heavier atoms, have remained undetected in metal lines with the exception of sodium detected in WASP-166b \citep{2022MNRAS.513L..15S,2020A&A...641L...7S} and a potential detection of \ion{C}{1} in HAT-P-11b \citep{2022NatAs...6..141B}.

Our AU Mic observations have a high enough signal-to-noise to look for transits of AU Mic b's escaping atmosphere in the host's metal emission lines. Figure~\ref{fig:metals_lc} shows the light curves during Visits 1 and 2 for \ion{N}{5}, \ion{O}{1}, \ion{Si}{4}, \ion{He}{2}, and \ion{C}{1}. All lines show decreasing flux around AU Mic b's mid-transit in Visit 1 except the slight emission shown by \ion{Si}{4}. It may be that the flare has a longer lasting impact on these metal lines as opposed to the one-orbit impact estimated for \lya\ and so they are still evolving towards quiescence during the planetary transit.

The Visit 2 \ion{N}{5}, \ion{O}{1}, \ion{He}{2} and \ion{C}{1} light curves have a similar shape to the Visit 2 \lya\ light curves in Figure~\ref{fig:lya_lc}, but with much lower signal-to-noise. \ion{N}{5} and \ion{He}{2}, two lines where we do not expect to see planetary material, exhibit the most pronounced transit-like behavior. The transit-like depth is reduced when the flux normalization is changed from the average between the fifth and sixth orbits to just the sixth orbit. The transit-like shape is also exhibited by the archival \ion{N}{5} and \ion{He}{2} light curves, which are well out of transit. As in \cite{2023AJ....165...12D}, we speculate that the metal line variability is due to some uncharacterized stochastic stellar activity potentially linked to magnetic heating processes within AU Mic's atmosphere. We encourage the use of this data in combination with future UV programs targeting M dwarfs to further investigate this non-flaring behavior.

The \ion{O}{1} and \ion{C}{1} light curves are of utmost interest when searching for planetary metal transits \citep{2022BAAS...54e2805O}. Although the \ion{O}{1} and \ion{C}{1} light curves in Figure~\ref{fig:metals_lc} depict transit-like behavior, we hesitate to characterize this as planetary absorption 1) because of the low signal from these lines and 2) because the behavior coincides with potential stellar activity signatures in the other emission lines presented. {\it HST}/COS will be more successful characterizing planetary metal line absorption than STIS. A metal outflow search is currently underway for AU Mic b using COS, which can be compared to these data to more definitively distinguish between planetary and stellar signatures (Feinstein, in prep).


\section{AU Mic b's High Energy Environment} \label{sec:highen}

We investigated the planet's high energy irradiation to place constraints on how much mass it is potentially losing and how much of it is staying neutral. We define high energy irradiation as the quiescent radiation received between $5 - 1170$ \AA, which covers the X-ray through the EUV. {\it ROSAT} observed AU Mic's quiescent X-ray luminosity to be $2.51\times 10^{29}$ erg s$^{-1}$, corresponding to a flux of $21500$ erg s$^{-1}$ cm$^{-2}$ at AU Mic b \citep[Table 3;][]{2015Icar..250..357C}.

Much of the star's EUV spectrum is inaccessible due to obscuration by the intervening interstellar medium. \cite{2021ApJ...913...40D}\footnote{\href{https://github.com/gmduvvuri/dem_euv}{https://github.com/gmduvvuri/dem\_euv}} modeled AU Mic's EUV spectrum using the differential emission measure technique. This method provides an estimate for the radiation from plasma at a specific temperature. They calculated an EUV flux of $8420$ erg s$^{-1}$ cm$^{-2}$ at AU Mic b. The combined XUV irradiation of the planet is about $29900$ erg s$^{-1}$ cm$^{-2}$.

\cite{2015Icar..250..357C} include X-ray and EUV luminosity measurements for a flaring AU Mic. Using the observed flare values from their Table 3, the combined flaring XUV irradiation of AU Mic b is $71500$ erg s$^{-1}$ cm$^{-2}$.

\subsection{Planetary mass-loss rate} \label{sec:massrate}

The amount of mass escaping the planet is often estimated by assuming the amount of high energy radiation received directly translates to heating the atmosphere, leading to a bulk flow of gas that overcomes the planet's gravitational potential. This is referred to as the energy-limited regime. There are two other photoevaporation regimes that may apply: recombination-limited and photon-limited \citep{2016ApJ...816...34O}. The flow timescale to recombination timescale ratio determines which regime applies to a planet. This ratio is largely governed by the planet's density and incident high energy flux. Figure $1$ in \cite{2016ApJ...816...34O} shows how the three photoevaporation regimes depend on a planet's mass and radius. Given AU Mic b's XUV irradiation of $29900$ erg s$^{-1}$ cm$^{-2}$, it will likely follow the top left or bottom right panels depending on the atmosphere's heating efficiency.

A typical source of uncertainty is the planet's mass. This is difficult to measure with extreme precision and is yet more difficult for planets with active hosts, like AU Mic, due to contamination from stellar variability. There are a few mass measurements reported for AU Mic b: $11.7\pm 5.0$ \Mearth \citep{2022MNRAS.512.3060Z}, $17.1^{+4.7}_{-4.5}$ \Mearth\ \citep{Klein2021} and $20.12^{+1.57}_{-1.72}$ \Mearth\ \citep{2021AJ....162..295C} from radial velocities, and $17\pm 5$ \Mearth\ from transit-timing variations \citep{2021AA...649A.177M}. 

AU Mic b has a radius of $2.60\times 10^{9}$ cm and a potential mass range of $6.99-12.0 \times 10^{28}$ g. With a heating efficiency of $0.1$ \citep[top left panel of Figure 1;][]{2016ApJ...816...34O}, AU Mic b will be within the energy-limited regime no matter its mass. However, if its atmosphere has a higher heating efficiency \citep[bottom right panel of Figure 1;][]{2016ApJ...816...34O}, then the planet could enter the photon-limited regime. During periods of high stellar activity, if a lower mass AU Mic b's XUV irradiation reaches $\gtrsim 10^5$ erg s$^{-1}$ cm$^{-2}$, then the planet could enter the recombination-limited regime. For this reason, we present AU Mic b's energy-limited mass loss rate as a possible scenario but the true mass loss rate could be smaller or larger by a few orders of magnitude and will vary on short and long timescales.

The energy-limited regime is given by Equation~\ref{eq:loss}, which is dependent on the incident high energy radiation ($F_{\text{XUV}}$), the planet's radius ($R_{\text{p}}$) and mass ($M_{\text{p}}$), the atmosphere's heating efficiency ($\eta$), and a correction factor ($K_{\text{eff}}$, Equation~\ref{eq:Keff}). $K_{\text{eff}}$ corrects for the difference in gravitational potential at the planet's radius versus its Roche lobe where the atmosphere needs to escape. 

\begin{equation} \label{eq:loss}
    \dot{M} = \eta \frac{\pi R^3_{\text{p}}F_{\text{XUV}}}{GM_{\text{p}}K_{\text{eff}}}
\end{equation}
where:
\begin{equation} \label{eq:Keff}
    K_{\text{eff}} = \frac{(a/R_{\text{p}} -1)^2 (2a/R_{\text{p}} +1)}{2(a/R_{\text{p}})^3}
\end{equation}

Assuming the properties listed in Table~\ref{tab:prop} and our quiescent XUV flux estimate, we estimate the energy-limited mass loss rate to be $(22.5-38.7)\eta \times 10^{10}$ g s$^{-1}$ for a mass range of $20.12-11.7$ \Mearth. We have not assigned a heating efficiency because it can vary widely across the exoplanet population \citep[0.1 - 0.6;][]{2014ApJ...795..132S}. The range presented is $100$ times larger than the AU Mic b mass loss rate estimated by \cite{2022AJ....164..110F}.

\subsection{Photoionization rate} \label{sec:photrate}

Radiation at wavelengths shorter than $912$ \AA\ can ionize neutral hydrogen in exoplanet atmospheres. Large fluxes at these energies have the ability to ionize much of the gas that is escaping a planet. The observability of the atmosphere at \lya\ depends on how much of the escaping material remains neutral for a long enough time to be accelerated to velocities that result in absorption in the line's blue- and red-wings. To get an idea for how significantly photoionization impacts AU Mic b, we estimated the photoionization rate with the following equation that quantifies the amount of ionization events per unit time ($\Gamma_{\text{ion}}$; count s$^{-1}$):

\begin{equation} \label{eq:photrate}
    \Gamma_{\text{ion}} = \int^{911.8\text{ \AA}}_{0\text{ \AA}} \frac{F_{\text{XUV}}(\lambda)\sigma_{\text{ion}}(\lambda)}{hc}\lambda d\lambda
\end{equation}
where $F_{\text{XUV}}$ is AU Mic b's wavelength-dependent XUV irradiation, and $\sigma_{\text{ion}}$ is the photoionization cross-section (cm$^2$) defined by
\begin{align*}
    \sigma_{\text{ion}} = &6.538\times 10^{-32} \left(\frac{29.62}{\sqrt{\lambda}}+1\right)^{-2.963} \\
    & \times (\lambda-28846.9)^2 \lambda^{2.0185}. \numberthis \label{eq:sigion}
\end{align*}

We split Equation~\ref{eq:photrate} into separate integrals over the X-ray ($5 < \lambda \leq 100$ \AA) and EUV ($100 < \lambda \leq 912$ \AA) spectral regions and summed them to get a photoionization rate of $4.01\times 10^{-4}$ s$^{-1}$. Inverting the photoionization rate characterizes the amount of time a typical neutral hydrogen atom stays neutral before interacting with an ionizing photon - the neutral hydrogen lifetime. A short neutral hydrogen lifetime indicates a smaller neutral atmosphere and likely a smaller \lya\ transit depth and shorter transit duration. 

At AU Mic b's orbital distance, the neutral hydrogen lifetime is about $0.692$ hours (almost $42$ minutes) during quiescence, and decreases by about $50\%$ during a flare. This value is short compared to the benchmark planet Gl 436b, which has a neutral hydrogen lifetime of 14 hours and consequently a large trailing neutral hydrogen tail \citep{2016A&A...591A.121B}. It is not entirely impossible to observe planets with short neutral hydrogen lifetimes, however, as shown by the detection of neutral hydrogen around GJ 3470b \citep[0.9 hours;][]{2018A&A...620A.147B}.

\cite{2023MNRAS.518.4357O} provide a theoretical framework that explains \lya\ detections and non-detections of atmospheric escape with photoionization and the host star's tidal influence. For example, the young Neptune K2-25b experiences enough XUV irradiation to have escaping material that is ionized and unobservable at \lya\, resulting in a repeatable non-detection \citep{2021AJ....162..116R}. A similar planet, HD 63433c, is less irradiated and is more likely to have escaping material detected at \lya\ \citep{2022AJ....163...68Z}. There has not yet been \lya\ observations of atmospheric escape on the same planet that vary to the degree we present in this work.

\subsection{Recombination rate}

If AU Mic b's neutral hydrogen outflow is dense enough, the photoionization from the Visit 1 flare -- and other instances of increased stellar emission -- would be overcome by the recombination of ionized hydrogen back into its neutral state. In this case, we would {\it not} expect the flare to cause the non-detection of a \lya\ transit in Visit 1. To test this, we estimated the recombination rate of AU Mic b's outflow and compared it to the flare photoionization rate from Section~\ref{sec:photrate}.

The neutral hydrogen column density of $10^{13.96}$ cm$^{-2}$ estimated in Section~\ref{sec:spec} is a lower limit since it only characterizes the fastest escaping material. For this reason, we use the mass loss rate calculated in Section~\ref{sec:massrate} and conservation of mass to estimate the total neutral hydrogen content of the outflow. Equation~\ref{eq:com} exhibits this relationship, with $\rho$ representing the density of neutral hydrogen in the outflow ($\rho = n_{\text{H I}}m_{\text{H I}}$), $r$ being some arbitrary distance from the planet, and $v$ as the outflow velocity ($20$ km s$^{-1}$ from Section~\ref{sec:lyalcurve}).

\begin{equation} \label{eq:com}
    \dot{M} = \rho (4\pi r^2)v
\end{equation}

Equation~\ref{eq:com} can be rearranged to solve for the number density of neutral hydrogen
\begin{equation} \label{eq:hdens}
    n_{\text{H I}} = \frac{\dot{M}}{(4\pi r^2)m_{\text{H I}}v}
\end{equation}
which gives an estimate of $10^{9.04}$ cm$^{-3}$ at the planet's white-light radius. The timescale for recombination is given by Equation~\ref{eq:recom}, where $\alpha$ is the recombination coefficient (assumed to be the Case A coefficient of $\alpha = 4.2 \times 10^{-13}$ cm$^3$ s$^{-1}$).

\begin{equation} \label{eq:recom}
    t_{\text{rec}} = (\alpha n_{\text{H I}})^{-1}
\end{equation}

Our resulting recombination timescale estimate is about $0.62$ hours. This is roughly double the neutral hydrogen lifetime due to a flare's photoionization rate from Section~\ref{sec:photrate}, which indicates that even the densest parts of AU Mic b's outflow will not have enough time to recombine during a flare to be visible in \lya\ transmission. The Visit 1 flare occurs about $6.6$ hours prior to AU Mic b's white-light transit and about $4.1$ hours prior to the planetary absorption seen in Visit 2. The $4.1$ hour difference between flare and potential neutral hydrogen tail transit is more than enough time for the neutral hydrogen to recombine. However, the \lya\ blue-wing mid-transit time from Section~\ref{sec:lyalcurve} is highly unconstrained since we do not observe the ingress of the planetary material in Visit 2. It is entirely possible that a leading neutral hydrogen outflow is ionized by the Visit 1 flare and thus rendered unobservable at \lya. Stellar activity and photoionization could play into our non-detection of a transit in Visit 1 versus a strong signal in Visit 2, which we discuss in Section~\ref{sec:discuss}.

\section{Summary \& Discussion} \label{sec:discuss}
\subsection{Visit 1}
The Visit 1 \lya\ spectra showed little to no evidence for absorption from intervening planetary material -- with the majority of its changes occurring in the red-wing. These red-wing flux changes can be seen in all three visits presented in Figure~\ref{fig:lyaspec} and do not depend on time away from planetary transit, leading us to believe they are caused by stellar activity. The Visit 1 metal line spectra showed similar amounts of stellar contamination, which more easily washes out planetary absorption at the lower line fluxes.

We characterized the flare observed in the first exposure of Visit 1. While the flare was energetic enough to hide any planetary absorption in \lya\ and other emission lines, its duration was estimated to be $1.5$ {\it HST} orbits -- much shorter than the expected planetary signal, and occurring $7$ hours before the white-light transit. 

Figure~\ref{fig:lya_lc} shows the Visit 1 \lya\ red-wing behavior is typical and consistent with the flux changes seen across all visits. The \lya\ blue-wing does exhibit some enhancement in Visit 1, likely attributable to stellar activity. The metal line Visit 1 light curves in Figure~\ref{fig:metals_lc} show decreasing flux behavior after the initial flare in the first orbit. We cannot positively identify or rule out a planetary absorption signature in the Visit 1 light curves, \lya\ or metals, because of our inability to effectively characterize the stellar contamination present. 

\subsection{Visit 2}
The Visit 2 \lya\ blue-wing in Figure~\ref{fig:lyaspec} features a stark increase in flux as opposed to the nominal red-wing flux changes. This blue-wing behavior is not present in either Visit 1 or the archival spectra. We attribute this behavior to intervening planetary neutral hydrogen that is transiting ahead of AU Mic b and is being accelerated away from the star. We reconstructed the ``out-of-transit" stellar \lya\ line using Visit 2's last orbit. This best-fit emission line profile and the subsequent best-fit ISM properties were combined with an additional absorption component to fit the ``in-transit" behavior of Visit 2's first orbit. We conclude that the planetary neutral hydrogen has a column density lower limit of $10^{13.96}$ cm$^{-2}$, and some of the material has been accelerated to a speed of $61.26$ km s$^{-1}$ radially away from the star.

The \lya\ light curves presented in Section~\ref{sec:lcurve} reinforce this picture. The \lya\ blue-wing light curve was modeled and fit with an opaque circular transiting object to characterize the size and timing of the escaping material. We find that the escaping neutral hydrogen is about $0.52$ $R_{\star}$ ($0.39$ \Rsol). This is just an estimate as we are limited by the geometry of our transit model, but this shows a large neutral hydrogen cloud is needed to explain the \lya\ blue-wing behavior. Following \cite{2023MNRAS.518.4357O}, we estimate the escaping material to be sculpted into a leading tail of length $1.39$ \Rsol\ and height $0.32$ \Rsol\ with an outflow velocity of $20$ km s$^{-1}$ before acceleration.

The Visit 2 metal line spectra and light curves do not show any significant behavior beyond what could be explained by stellar activity. While the \ion{N}{5} and \ion{He}{2} Visit 2 light curves in Figure~\ref{fig:metals_lc} look similar to the \lya\ blue-wing light curve in Figure~\ref{fig:lya_lc}, this could be because of a flux increase later in the visit. The second and third orbits within the \ion{Si}{4} light curve do show some flux variations that may be indicative of stellar activity occurring throughout the visit. \ion{O}{1} and \ion{C}{1} are too low signal-to-noise to detect transiting planetary material.

\subsection{Bigger picture}

While no planetary absorption could be definitively identified in Visit 1, Visit 2 indicates planetary neutral hydrogen is escaping ahead of AU Mic b and being accelerated away from the host star. \cite{2017A&A...605L...7L} reported slight changes in the shape of Gl 436b's escaping atmosphere shown by its changing \lya\ transit shape across eight epochs, but detection of the outflow is repeatable. Our work on AU Mic b is the first time a variable atmospheric escape signature in \lya\ has been observed to greater magnitude -- going from undetectable to detectable.

\cite{2019ApJ...873...89M} used 3D hydrodynamic modeling of planetary outflows within varying stellar wind environments to show the geometry of the outflow changes with stellar wind strength. They predicted that an intermediate stellar wind strength could shape a time-variable ``dayside arm" of planetary material. The intermediate strength stellar wind balances against the planetary outflow, creating fluid instabilities that result in a growing dayside arm that eventually detaches from the planet \citep[Figure 12,][]{2019ApJ...873...89M}. This ``burping" scenario could explain our observation of \lya\ absorption ahead of AU Mic b's transit in Visit 2 and its time-dependence also explains why we do not see \lya\ absorption in Visit 1. If this is the case, this work presents the first observational evidence of the burping behavior predicted by \cite{2019ApJ...873...89M} and affirms the power of \lya\ planet transits in characterizing stellar wind environments.

The potential intermediate stellar wind strength argued above is at odds with the extreme stellar wind strength scenarios depicted in \cite{2020MNRAS.498L..53C} and \cite{2022ApJ...934..189C}. \cite{2020MNRAS.498L..53C}, motivated by the unknown stellar wind environment of exoplanets orbiting M dwarfs, simulated AU Mic b's outflow within different stellar wind strengths. Using 3D hydrodynamics, they varied AU Mic's mass loss rate and found that AU Mic b's \lya\ transit could be nearly undetectable in the most extreme case, a stellar mass loss rate of $1000$ $\dot{\text{M}}_{\odot}$. \cite{2022ApJ...934..189C} used a global magnetosphere magnetohydrodynamic code \citep[BATS-R-US;][]{2012JCoPh.231..870T,1999JCoPh.154..284P} to simulate AU Mic b's escaping neutral hydrogen atmosphere within a time-varying stellar wind environment and output corresponding \lya\ absorption light curves. The atmosphere's response varied over the course of the planet's orbit, which contributed to the planet's \lya\ light curves changing shape from transit to transit. This builds off of the work done by \cite{2021ApJ...913..130H} to model the magnetohydrodynamic interactions between a planetary outflow and varying stellar wind conditions. They found that planetary \lya\ transit signatures could vary on short (hour-long) timescales due to the outflow actively changing shape throughout the orbit under its local stellar wind conditions (see their Figure 6). These works show that magnetized stellar wind strength, at least in part, could explain the difference we see between our AU Mic b \lya\ light curves although neither present an explanation for the \lya\ absorption occurring ahead of the white-light transit.

Photoionization is another major factor in the observability of planetary neutral hydrogen. The short \lya\ transit duration of GJ 3470b is caused by the quick photoionization of its outflow \citep{2018A&A...620A.147B}, and the inability to detect K2-25b and HD 63433b at \lya\ is thought to be caused by the near total photoionization of their outflows \citep{2023MNRAS.518.4357O}. Given AU Mic's penchant for flaring, AU Mic b could be experiencing variable amounts of ionizing radiation. We showed in Section~\ref{sec:highen} that neutral hydrogen could be photoionized within 42 minutes of escaping AU Mic b, before being accelerated to observable speeds, just from AU Mic's quiescent radiation. This is consistent with the short, yet largely unconstrained, \lya\ transit duration we observe in Visit 2. During a flare, the neutral hydrogen lifetime could become as short as 22 minutes and further impact the observability of a \lya\ transit. We do not know to what extent self-shielding from photoionization and radiation pressure are also involved \citep{2015A&A...582A..65B,2013A&A...557A.124B}.

The data presented in this work can tune future FUV observation plans to better sample AU Mic b's transit -- probing before the white-light transit and further constraining its \lya\ transit shape. These observations would help answer questions about the timing of AU Mic b's \lya\ transit, how frequent and to what magnitude the transit varies, the impact of stellar wind and flares, and observing the escape of metals. Its drastic change in \lya\ light curve behavior makes AU Mic b a prime candidate for continued characterization of its environment and modeling of its atmosphere. We could distinguish between the potential stellar wind and flare scenarios posed above by confirming the repeatability of the \lya\ transit ahead of the planet. Applying increasingly sophisticated simulations to this system will help us learn more about the extreme behavior of close-in planets around young stars.

\acknowledgments
We sincerely thank Dr. Hans R. M\"uller, Dr. Brian Chaboyer, and Dr. James Owen for their help and guidance throughout the progress of this work. The authors appreciate the support, care, and community provided by the graduate students within Dartmouth College's Department of Physics \& Astronomy. We would like to express our sincere appreciation for Nova, Jacques, Margot, Charlie, Edmund, and Nessie who have begrudgingly accepted our transition back to in-person work. We will miss you, Charlie. Welcome to the family Jonah and Theo!

This research is based on observations made with the NASA/ESA {\it Hubble Space Telescope} obtained from the Space Telescope Science Institute, which is operated by the Association of Universities for Research in Astronomy, Inc., under NASA contract NAS 5–26555. These observations are associated with HST-GO-15836. Support for program HST-GO-15836 was provided by NASA through a grant from the STScI.

CHIANTI is a collaborative project involving George Mason University, the University of Michigan (USA), University of Cambridge (UK) and NASA Goddard Space Flight Center (USA).

\appendix
\section{Additional spectrum information}
\subsection{Far-Ultraviolet Continuum} \label{ap:cont}
The FUV continuum observed by {\it HST}/STIS using the E140M grating was defined as a combination of the overlapping continuum presented in \cite{2022AJ....164..110F} and low-signal regions identified by eye: $1152.602 - 1155.579$, $1159.276 - 1163.222$, $1164.565 - 1173.959$, $1178.669 - 1188.363$, $1195.162 - 1196.864$, $1201.748 - 1203.862$, $1227.056 - 1236.921$, $1262.399 - 1263.967$, $1268.559 - 1273.974$, $1281.396 - 1287.493$, $1290.494 - 1293.803$, $1307.064 - 1308.703$, $1319.494 - 1322.910$, $1330.349 - 1332.884$, $1337.703 - 1341.813$, $1341.116 - 1350.847$, $1356.5 - 1370.5$, $1372.5 - 1392.0$, $1395.2 - 1401.0$, $1404.5 - 1521.5$, $1522.5 - 1532.5$, $1535.0 - 1545.0$, $1553.0 - 1560.0$, $1562.5 - 1638.5$, $1642.5 - 1655.5$, $1658.0 - 1670.0$, $1672.5 - 1700.0$ \AA.

\begin{figure*}[b!]
    \centering
    \includegraphics[width=\textwidth]{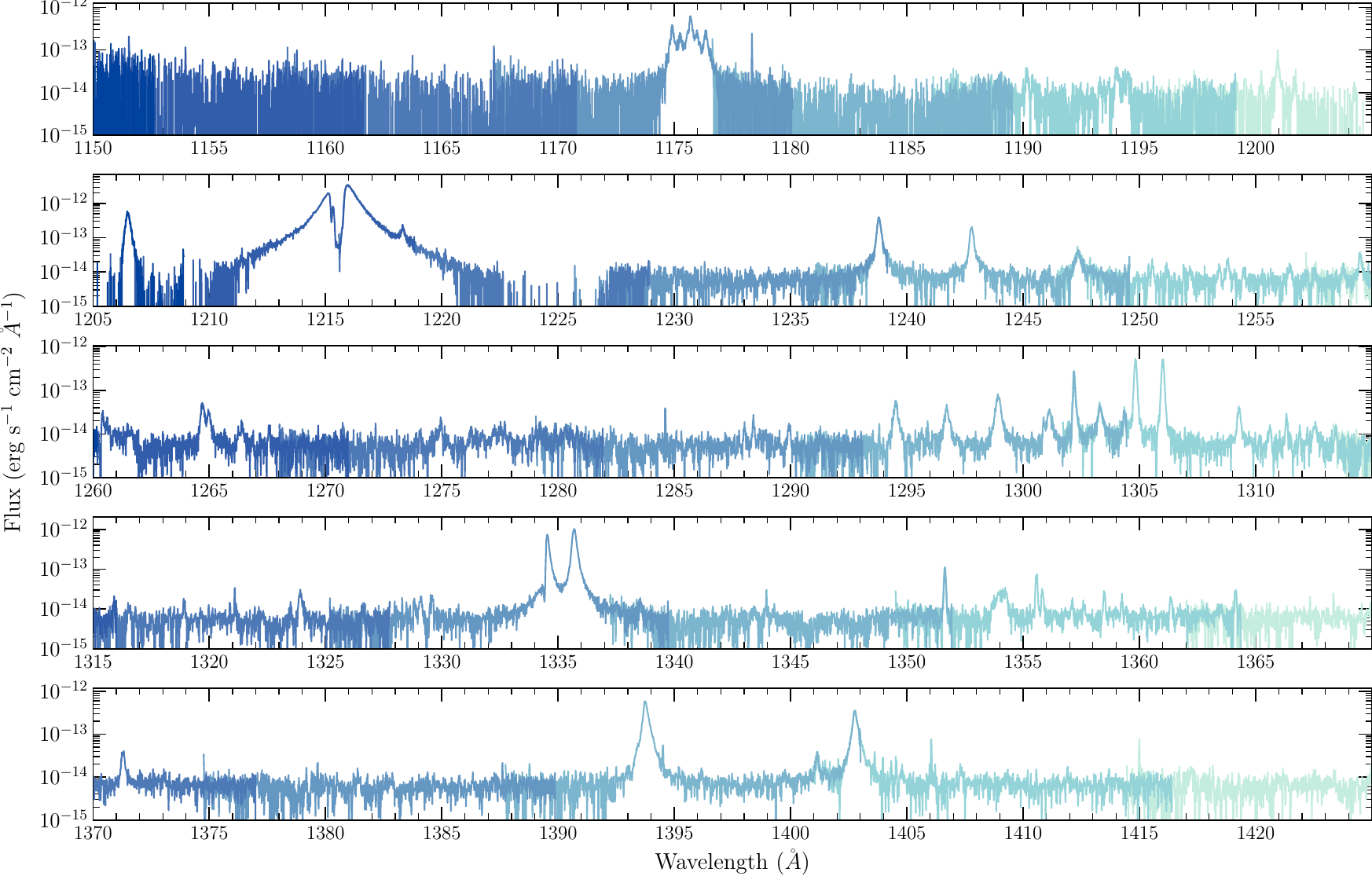}
    \caption{AU Mic's quiescent spectrum -- the last orbit from Visit 1. Continued in Figure~\ref{fig:full_spec2}.}
    \label{fig:full_spec1}
\end{figure*}

\begin{figure*}
    \centering
    \includegraphics[width=\textwidth]{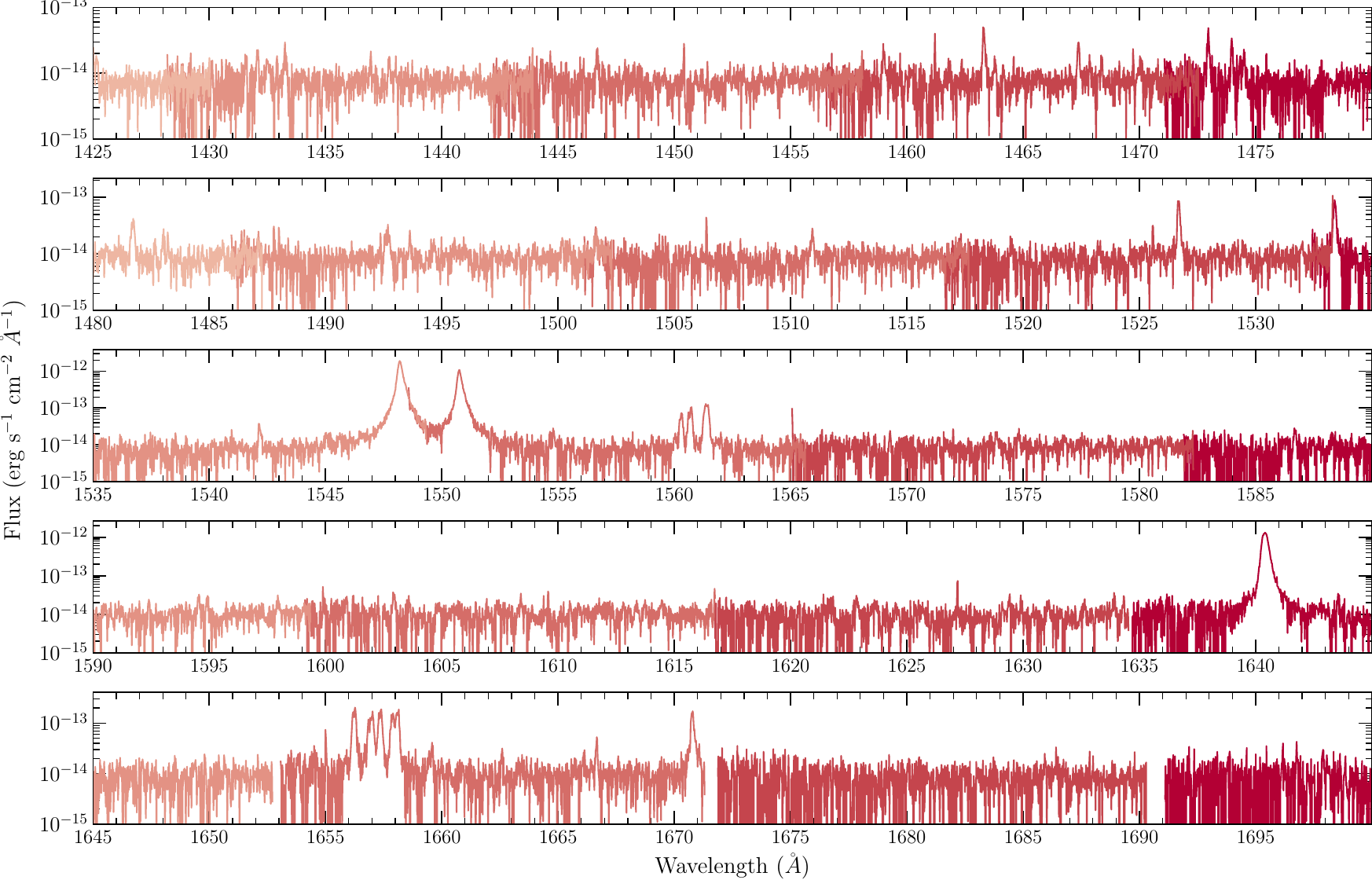}
    \caption{AU Mic's quiescent spectrum -- the last orbit from Visit 1.}
    \label{fig:full_spec2}
\end{figure*}

\vspace{5mm}
\facility{{\it HST}/STIS}

\software{
        {\tt astropy} \citep{2018AJ....156..123T},~
        {\tt batman} \citep{2015PASP..127.1161K},~
        {\tt cos\_flares} \citep{2022AJ....164..110F},~
        {\tt emcee} \citep{2013PASP..125..306F},~
        {\tt lightkurve} \citep{2018ascl.soft12013L},~
        {\tt lyapy} \citep{2016ApJ...824..101Y},~
        {\tt matplotlib} \citep{hunter2007matplotlib},~
        {\tt numpy} \citep{harris2020array},~
        {\tt scipy} \citep{2020SciPy},~
        {\tt stistools} (\url{https://github.com/spacetelescope/stistools})
        }

\newpage
\bibliography{references}

\end{document}